\documentclass[12pt,letterpaper]{article}
\usepackage[toc,title,page]{appendix}

\usepackage{amssymb}
\usepackage{amsmath}
\usepackage{amsfonts}
\usepackage{amsthm}
\usepackage{multirow}
\usepackage{color}
\usepackage{url}
\usepackage[pdftex]{graphicx}
\usepackage[margin=1in]{geometry}

 \newtheorem{lem}{Lemma}

 \newtheorem{prop}{Proposition}
 \newtheorem{cor}{Corollary}

 \theoremstyle{definition}
 \newtheorem{exa}{Example}

\begin{document}

\title{\textsc{Raising Bidders' Awareness in Second-Price Auctions}\thanks{We thank the late Nora Szech for initial discussions on the project that led to the paper. We thank the editors and two anonymous reviewers for their helpful comments. Ying Xue gratefully acknowledges financial support from the National Natural Science Foundation of China ($\sharp$72003147). Burkhard gratefully acknowledges financial support via ARO Contract W911NF2210282.}}

\author{
  Ying Xue Li\thanks
  {Jinhe Center for Economic Research, Xi'an Jiaotong University, and School of Economics and Management, Xinjiang University, liyxjinhe@xjtu.edu.cn}
  \and
  Burkhard C. Schipper\thanks
  {Department of Economics, University of California, Davis, bcschipper@ucdavis.edu}
}

\date{September 12, 2026}

\maketitle

\begin{abstract} When bidders bid on complex objects, they might be unaware of characteristics affecting their valuations. We assume that each buyer's valuation is a sum of independent random variables, one for each characteristic. When a bidder is unaware of a characteristic, he omits the random variable from the sum. We study the seller's decision to raise bidders' awareness of characteristics before a second-price auction with entry fees. Optimal entry fees capture an additional unawareness rent due to unaware bidders misperceiving their probability of winning and the price to be paid upon winning. When raising a bidder's individual awareness of a characteristic with positive expected value, the seller faces a trade-off between positive effects on the expected first order statistic and unawareness rents of remaining unaware bidders on one hand and the loss of the unawareness rent from the newly aware bidder on the other. We present results on raising public awareness together with no versus full information. We discuss the winner's curse due to unawareness of characteristics. 
\newline
\newline
\noindent \textbf{JEL-Classification: } C72, D44, D83. 
\end{abstract}

\newpage

\rightline{\textit{In memory of Nora Szech.}}

\section{Introduction}

We study raising bidders' awareness before second-price auctions with independent private values. An auctioneer wants to sell an object to one of $n$ risk-neutral bidders. The bidders' valuations for the object depend on multiple characteristics, some of which bidders might be even unaware of. That is, a bidder may not conceive of all characteristics of the object that may affect his valuation. Such situations are common when the object is for instance an ``experience good'' whose characteristics become only transparent when it is used by the future winning bidder. In such contexts, it is also natural to assume that the auctioneer, as a prior owner, has more experience with the object and is already aware of the relevant characteristics. Consequently, the auctioneer may now decide to strategically raise the bidders' awareness of some characteristics but not of others, and may also commit to disclose some amount of information on these characteristics. We emphasize that raising awareness of a characteristic just means telling the bidder about the existence of the characteristic. This is different from providing information about the value of the characteristic. The latter implies the former but not vice versa. To emphasize the difference between awareness and information, consider a statement like ``the neighbor of the property may or may not have a prescriptive easement''. Such a statement surely raises awareness of the characteristic ``prescriptive easement'' but does not provide information about the likelihood of such an easement because the statement is a tautology, i.e., an event that always obtains. Similarly, a statement such as ``the neighbor has a prescriptive easement and does not have a prescriptive easement'' raises awareness of such an easement in the sense of letting the recipient of the statement think about it but it does not provide information about it because it is a contradiction, i.e., an event that never obtains. Without awareness of a characteristic, the bidder cannot form beliefs about the value of the characteristic. In contrast, disclosure of information means that also some information is provided about the value of the characteristic like for instance ``it is more likely than not that the neighbor has a prescriptive easement on the property''. Such a statement not only raises awareness of a prescriptive easement but also provides information on it.  

Our work is inspired by Szech (2011). She studied the (possibly asymmetric) disclosure of costly information before second-price auctions with entry fees in a setting of independent private values and risk neutral bidders. We generalize her characterization of optimal entry fees (her Proposition 1) to settings that allow the seller to potentially raise asymmetric awareness (see our Proposition~\ref{revenue} and Corollary~\ref{Szech}). Yet, while she then studies the surprising optimality of asymmetric disclosure of costly information to bidders, we focus on costless raising awareness and costless disclosure of information. 

Before describing our approach and results as well as the connection to the rest of the literature, we like to mention as a motivation three potential applications: 

\noindent \textbf{Firms competing in an auction to take over a firm: } The owner of the firm, as an insider, has not just better information on the firm but is also aware of all hidden details of the business of the firm including potential law suits, innovative products, potential accounting problems, etc. while some of the bidders may not even think about them. The owner has to decide of what she raises the bidders' awareness and how much information she provides beyond mandatory disclosure requirements. A similar motivation, but just w.r.t. information, has been used by Bergemann and Pesendorfer (2007), Ganuza (2004), Eso and Szentes (2007), and Szech (2011) among others, and seems to be based on Burkart (1995). 

\noindent \textbf{Bidding for access to internet users: } Advertisers compete in an auction for displaying advertisement to users of websites, search engines, and apps. Internet firms such as Google, Facebook etc. collect massive amounts of data on individual users. Big data about users allow for inferences of detailed psycho-demographic profiles of internet users (Kosinski, Stillwell, and Graepel, 2013). These profiles are of potential interest to advertisers for estimating their willingness to pay for an ad shown to the user. Yet, advertisers may not be aware of all dimensions on which data is collected and some of the information may not be provided to advertisers either for strategic reasons or because of privacy laws. The internet firms still can decide about what to tell the advertisers about the characteristics on which data is collected on, even if the data cannot be provided to advertisers.\footnote{In a recent paper, Gao (2025) studies disclosure of information via big data sets (not in an auction though). She considers selective disclosure of information only, which we can think of represented by the ``rows'' of a data set (i.e., the individual observations). In contrast, we think of raising awareness of ``columns'' of a data set, the characteristics on which data is collected.} 

\noindent \textbf{Procurement/Sale for complex projects/commodities: } Potential contractors compete for an award of a complex project/commodity such as for a novel weapon technology, a power plant, a property etc. These projects involve many dimensions relevant to quality, costs, values etc. Contractors may not be aware of some of those dimensions as they are not necessarily transparent to outsiders. In contrast, the project owner may be aware of those dimensions and may have information on those dimensions. Before running the procurement or sales auction, the owner can decide about the issues she wants to make bidders aware and commit on the information she provides on these issues. 

To model unawareness of characteristics relevant to the bidders' valuations, we allow in Section~\ref{model} for multi-dimensional valuations in the form of random vectors with each component of the vector representing the value of one characteristic. A bidder's valuation is now the sum of the component random variables. Yet, the bidder may be unaware of some of those dimensions, only summing over the random variables of which he is aware. Being unaware of a characteristic and the random variable representing this characteristic also means that the bidder does not consider that other bidders may be aware of it. Even though we use the weak dominant strategy equilibrium of independent private-value second-price auctions, these beliefs about other's awareness are relevant because we are interested in optimal entry fees that the auctioneer can charge, which depend on the bidders' expected profit. We model payoff type spaces with unawareness as a restricted version of unawareness type spaces by Heifetz, Meier, and Schipper (2013a). Our model comprises a finite lattice of payoff type spaces with projections, one payoff type space for every subset of characteristics (that includes a ``default'' characteristic) modeling differences in awareness. The lattice order is naturally induced by set inclusion on the subsets of characteristics. Information is modeled via sigma-algebras on those payoff type spaces. When the seller discloses a sigma-algebra to a bidder, the payoff type space on which the sigma-algebra is defined represents the awareness level and the elements of the sigma-algebra represent the information. 

Our first result in Section~\ref{entry_fees}, Proposition~\ref{revenue}, characterizes the auctioneer's optimal entry fees and expected revenue from the weak dominant strategy equilibrium of the second-price auction. Compared to Szech (2011), we obtain an additional term in the auctioneer's expected revenue that for each bidder captures the difference of the bidder's expected profit and the bidder's expected profit from the perspective of someone with full awareness. This unawareness rent is only present when the bidder is unaware of a characteristic and bidders have asymmetric awareness. It is caused by the bidder's misperception of his probability of winning and his misperception of the expected price he has to pay upon winning. 

Armed with the characterization of optimal entry fees given awareness and information, we study in Section~\ref{individual_awareness} how to optimally raise individual bidders' awareness. We show that if a characteristic has positive expected value, then the auctioneer wants to make at least one bidder aware of it, creating an asymmetry in awareness and thus unawareness rents in addition to increasing the expected (largest) first order statistic. Raising further bidders' awareness involves now a trade-off: It increases further the expected first order statistic. It also increases further the unawareness rents accrued from remaining unaware bidders (because they overestimate their probability of winning and underestimate the expected price they have to pay upon winning even more). However, the auctioneer also loses now the unawareness rent from the bidder whom she made aware of the characteristic. We show by examples that a characteristic having positive expected value is not necessary for raising awareness of the characteristic. 

We also consider optimally raising public awareness, when any awareness disclosed must be disclosed to all bidders. We show that the auctioneer raises public awareness of a characteristic with no information if and only if the expected value of the characteristic is positive. When raising awareness involves mandatory disclosure of full information, the auctioneer keeps the public unaware of a characteristic whenever the characteristic contributes negatively to the first order statistic, and only if the expected value of the characteristic is negative. 

In Section~\ref{winners_curse}, we discuss the winner's curse due to unawareness of characteristics. When a seller optimally raises awareness, then any characteristics of which bidders are kept unaware must have negative expected value. Thus, unaware winners pay more in expectation than what they would if they were aware of all characteristics. This winner's curse due to unawareness occurs even with independent private values and is therefore different from the winner's curse in common value auctions. 

In Section~\ref{awareness_vs_information}, we discuss how awareness differs from information. We show that given common awareness of a set of characteristics, the auctioneer wants to optimally disclose all information on these characteristics. This generalizes an observation by Szech (2011), who noted that in the absence of information costs the auctioneer wants to disclose all information because of the convexity of the expected maximum. It is in contrast to Ganuza (2004), who finds that the auctioneer does not always want to disclose all information before second-price auctions. However, he does not allow for entry fees (see also Board, 2009, Hagedorn, 2009, Bergemann et al., 2022). Our result is reminiscent of Eso and Szentes (2007) who also show the optimality of full disclosure but in a more general mechanism design setting (see also Gershkov, 2009).\footnote{Other papers on information disclosure before auctions include Milgrom and Weber (1982), Rasmusen (2006), Bergemann and Pesendorfer (2007), Vagstad (2007), Ganuza and Penalva (2010, 2019), and Arefeva and Meng (2021).}

A discussion of the related literature on contracting under unawareness is contained in Section~\ref{unaw_lit}. Further mathematical details that should be useful in future applications are collected in Appendix~\ref{additional}. Proofs are relegated to Appendix~\ref{proofs}.

\section{Varying Dimensions of Characteristics\label{model}} 

Consider a seller who wants to sell one indivisible object to one of $N = \{1, ..., n\}$ risk-neutral buyers via a second-price auction with entry fees. Each buyer's willingness to pay for the object depends on the values of characteristics of the object. There are up to $m$ characteristics, $M = \{1, ..., m\}$. The values of characteristics are coded into a $m$-dimensional real-valued vector of variables. That is, each characteristic is represented by one dimension of the vector. 

Bidders do not know the value of each characteristic nor are they necessarily aware of all characteristics. Before the auction, the seller can decide for each bidder $i \in N$ of which characteristics $M^i \subseteq M$ she wants to make him aware. Moreover, she can also decide on how much information the bidder can learn about the value of these characteristics. This requires us to consider second-price auctions with multi-dimensional valuations in which the dimension may vary with the disclosure decision by the seller.

Define $\mathcal{M} = \{M' \in 2^M : 1 \in M' \}$. This is the set of all subsets of characteristics that contain the first characteristic. Since $M$ is finite, we have that $\mathcal{M}$ is finite. The reason for considering only subsets of characteristics that contain the first characteristic is that we want valuations to be always well-defined. That is, all bidders are always aware of the first characteristic. For each $M' \in \mathcal{M}$, we let $S_{M'}$ be a measurable space of states with sigma-algebra $\mathcal{F}_{M'}$. For any $M', M'' \in \mathcal{M}$, $M' \neq M''$, we let $S_{M'}$ and $S_{M''}$ be disjoint. The set of all such spaces is denoted by $\mathcal{S} = \{S_{M'}\}_{M' \in \mathcal{M}}$. Observe that $\mathcal{S}$ is a finite lattice of disjoint spaces with the lattice order induced by set inclusion of subsets in $\mathcal{M}$ with the unique greatest element $S_{M}$ and the unique least element $S_{\{1\}}$. We let $\Omega := \bigcup_{M' \in \mathcal{M}} S_{M'}$ be the (disjoint) union of spaces. For every $M'', M' \in \mathcal{M}$ with $M' \subseteq M''$, there is a measurable surjective projection $r^{M''}_{M'}: S_{M''} \longrightarrow S_{M'}$ that satisfies (i) $r^{M'''}_{M'} = r^{M''}_{M'} \circ r^{M'''}_{M''}$ for every $M''', M'', M' \in \mathcal{M}$ with $M' \subseteq M'' \subseteq M'''$, and (ii) $r^{M'}_{M'} = id_{M'}$ (i.e., the identity on $S_{M'}$) for every $M' \in \mathcal{M}$. These projections allow us to model events across spaces. The lattice structure is a special case of unawareness structures by Heifetz, Meier, and Schipper (2013a). It enables us to model situations in which the seller keeps some bidders unaware of some characteristics of the object. 

For every characteristic $j \in M$ and bidder $i \in N$, we introduce a measurable random variable $X_j^i: S_{\{1, j\}} \longrightarrow \mathbb{R}$. We assume that $X_j^i$ is almost surely not constant on $S_{\{1, j\}}$ and integrable. We extend $X_j^i$ to states in spaces in $\{S_{M'} \in \mathcal{S} : j \in M', M' \in \mathcal{M}\}$ by letting for $M' \in \mathcal{M}$ with $j \in M'$ and $\omega \in S_{\{1, j\}}$, $X_j^i(\omega') = X_j^i(\omega)$ for all $\omega' \in (r^{M'}_{\{1, j\}})^{-1}(\omega)$.  That is, the random variable on $S_{M'}$ is extended to more expressive spaces using the inverse image with respect to the measurable surjective projections. For every $M' \in \mathcal{M}$, we introduce a measurable random vector $X_{M'}: \bigcup_{S \in \{S_{M''} \in \mathcal{S}: M' \subseteq M'', M'' \in \mathcal{M}\}} S \longrightarrow \mathbb{R}^{n \cdot |M'|}$ by $X_{M'}(\omega) = (X_j^i(\omega))_{j \in M', i \in N}$ for all $\omega \in \bigcup_{S \in \{S_{M''} \in \mathcal{S}: M' \subseteq M'', M'' \in \mathcal{M}\}} S$. Lemma~\ref{measurableX} in Appendix~\ref{additional} allows agents with awareness higher than $M'$ to reason about random vectors $X_{M'}$. 
\begin{figure}\caption{Illustration of an Unawareness Structure\label{illustration}}
\begin{center}
\includegraphics[scale=.1]{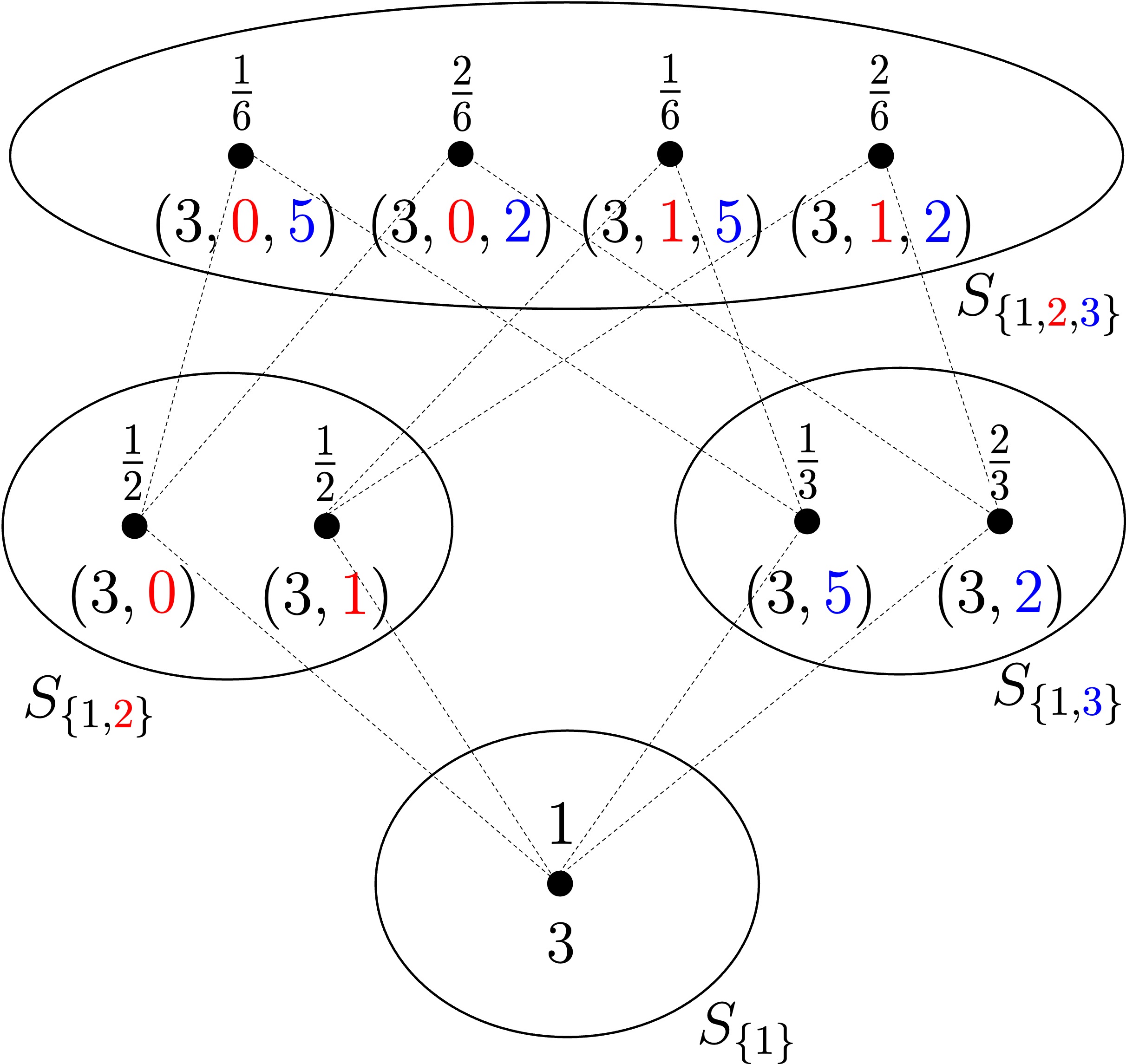}
\end{center}
\end{figure}

For any $M' \in \mathcal{M}$, let $P^{M'} \in \Delta(S_{M'})$ denote a probability measure on $S_{M'}$. For any $M'', M' \in \mathcal{M}$ with $M' \subseteq M''$, the marginal of $P^{M''}$ on $S_{M'}$, denoted by $P^{M''}_{|M'}$, is defined by $P^{M''}_{| M'}(D) : = P^{M''}\left((r_{M'}^{M''})^{-1}(D)\right)$ for $D \in \mathcal{F}_{M'}$. Let $P = (P^{M'})_{M' \in \mathcal{M}} \in \times_{M' \in \mathcal{M}} \Delta(S_{M'})$ be a projective system of probability measures. I.e., $P^{M'}$ is the marginal of $P^{M''}$, for every $M', M'' \in \mathcal{M}$ with $M' \subseteq M''$.\\

Figure~\ref{illustration} illustrates the features of our setting introduced so far in a simplified finite example. There are three characteristics $M = \{1, \textcolor{red}{2}, \textcolor{blue}{3}\}$. Hence, $\mathcal{M} = \{\{1\}, \{1, \textcolor{red}{2}\}, \{1, \textcolor{blue}{3}\}, \\ \{1, \textcolor{red}{2}, \textcolor{blue}{3}\}\}$. There are four spaces, $S_{\{1\}}, S_{\{1, \textcolor{red}{2}\}}, S_{\{1, \textcolor{blue}{3}\}}$, and $S_{\{1, \textcolor{red}{2}, \textcolor{blue}{3}\}}$ that are partially ordered by set inclusion on $M$. That is, both $S_{\{1, \textcolor{red}{2}\}}$ and $S_{\{1, \textcolor{blue}{3}\}}$ are richer than $S_{\{1\}}$ and poorer than $S_{\{1, \textcolor{red}{2}, \textcolor{blue}{3}\}}$. Among each other, $S_{\{1, \textcolor{red}{2}\}}$ and $S_{\{1, \textcolor{blue}{3}\}}$ are incomparable. The spaces are depicted in order in Figure~\ref{illustration}. As mentioned above, we illustrate our setting with a finite example. So we only indicated by solid bullets a few states in each space. The dash lines between states across spaces indicate the surjective projections from richer to poorer spaces. For simplicity, we consider just one bidder and therefore omit the index of the bidder. The number(s) \emph{below} each bullet belong to a valuation random variable. The random variable $X_1$ is defined on $S_{\{1\}}$ and extended to all spaces by assigning the same number to all states in the inverse image of the state in the lowest space. That is, $X_1$ is constant across all states, assigning the number 3 (the first component of the vectors in spaces richer than $S_{\{1\}}$). Similarly, the random vector $X_{\{1, \textcolor{red}{2}\}}$ is defined on $S_{\{1, \textcolor{red}{2}\}}$ (coinciding of course with the random variable $X_{\{1\}}$ in the first component) and extended to the richer space $S_{\{1, \textcolor{red}{2}, \textcolor{blue}{3}\}}$ by requiring the same value for the inverse images. To make the characteristics easily distinguishable in the illustration, we use red to indicate the second characteristic and blue to indicate the third characteristic in Figure~\ref{illustration}. Finally, the system of projective probability measures $P$ is indicated by the probabilities \emph{above} each state in Figure~\ref{illustration}. We observe that the probability on $S_{\{1, \textcolor{red}{2}\}}$ is the marginal of the probability measure on $S_{\{1, \textcolor{red}{2}, \textcolor{blue}{3}\}}$, etc., so that indeed the probabilities on each states form a system of four probability measures.\\ 

Because $X_{M'}$ is measurable, for any $M' \in \mathcal{M}$, we have by Lemma~\ref{measurableX} in Appendix~\ref{additional} that for any $x_{M'} \in \mathbb{R}^{n \cdot |M'|}$, that the set $\{\omega \in \Omega : X_{M'}(\omega) \leq x_{M'}\} \cap S_{M''} \in \mathcal{F}_{M''}$ for any $M'' \in \mathcal{M}$ with $M' \subseteq M''$. We define $$P^{M''}\left(\{\omega \in \Omega : X_{M'}(\omega) \leq x_{M'}\}\right) := P^{M''}\left(\{\omega \in \Omega : X_{M'}(\omega) \leq x_{M'}\} \cap S_{M''}\right)$$ if $M'' \in \mathcal{M}$ such that $M' \subseteq M''$ and let it be undefined otherwise.

In light of Lemma~\ref{P} in Appendix~\ref{additional}, we abuse notation and write $P(\{\omega \in \Omega : X_{M'}(\omega) \leq x_{M'}\})$ for $P^{M''}\left(\left\{\omega \in \Omega : X_{M'}(\omega) \leq x_{M'}\right\} \right)$ for any $M'' \in \mathcal{M}$ with $M' \subseteq M''$. Moreover, we can now define the distribution function of the random vector $X_M$ by $$F(x_M) = P(\{\omega \in \Omega : X_M(\omega) \leq x_M\}).$$ Similarly, for any subset of characteristics $M' \in \mathcal{M}$ and nonempty subset of bidders $N' \subseteq N$, we let $F_{M'}^{N'}(x_{M'}^{N'})$ be the marginal distribution, where $x_{M'}^{N'} = (x_j^i)_{j \in M', i \in N'} \in \mathbb{R}^{|N'| \cdot |M'|}$. If no confusion arises, we abuse notation and write $F(x_{M'}^{N'})$ for the marginal distribution.

We adopt the common simplifying assumption in the literature on standard auctions that values are drawn independently across bidders. We will also assume that values of characteristics are drawn independently. The set of random variables $\{X_j^i\}_{j \in M, i \in N}$ is mutually independent. For every $M' \in \mathcal{M}$, every vector $(x_j^i)_{j \in M', i \in N} \in \mathbb{R}^{n \cdot |M'|}$, and every nonempty $N_j \subseteq N$ (where $N_j$ may be interpreted as the subset of bidders who are aware of characteristic $j$),
\begin{eqnarray*} P\left( \bigcap_{j \in M'} \bigcap_{i \in N_j}\{\omega \in \Omega : X_j^i(\omega) \leq x_j^i \} \right) = \prod_{j \in M'} \prod_{i \in N_j} P \left(\{\omega \in \Omega : X_j^i(\omega) \leq x_j^i\}\right).
\end{eqnarray*}

In terms of distributions, we have that for any $M' \in \mathcal{M}$, $j \in M'$, $\emptyset \neq N_j \subseteq N$, and $x_j^{N_j} = (x_j^i)_{i \in N_j}$,
\begin{eqnarray*} F((x_j^{N_j})_{j \in M'}) = \prod_{j \in M'} \prod_{i \in N_j} F_j^i(x_j^i).
\end{eqnarray*}
We assume that individual marginal distributions for a characteristic are commonly known among all bidders who are aware of the characteristic.

\section{Optimal Entry Fees given Disclosure of Awareness and Information\label{entry_fees}} 

Recall that bidders may not be aware of all characteristics. Initially, they are all aware of characteristic $1$ only. Before the auction, the seller may make bidders aware of further characteristics and allow bidders to learn some information on the value of these characteristics. In return, she may charge each bidder an individual participation fee and commits to exclude any bidder from the auction who does not pay the participation fee. While we assume that the seller is aware of all characteristics, the seller does not know the precise information that the bidder learns after he is charged the participation fee. Formally, the seller describes to each bidder $i$ a sigma-algebra $\mathcal{I}^i_{M'} \subseteq \mathcal{F}_{M'}$ for characteristics $M' \in \mathcal{M}$ on which bidder $i$ will receive information. Note that the seller's description of this sigma-algebra makes the bidder already aware of characteristics in $M'$. If $M' \in \mathcal{M}$ is the set of characteristics of which bidder $i$ is made aware of by the seller, then we denote $M^i = M'$. We call $M^i$ bidder $i$'s awareness level. The process of information disclosure is inspired by Szech (2011) except that in her model all bidders are aware of all characteristics upfront.

Any bidder may form beliefs about the awareness of other bidders including bidders whom the seller made aware of larger or incomparable subsets of characteristics. We will assume that bidder $i$ forms correct beliefs about the awareness of other bidders subject to his own awareness. That is, bidder $i$ cannot believe that another bidder is aware of a specific characteristic that he himself is unaware of. Formally, we assume that if $(\mathcal{I}^i_{M^i})_{i \in N}$ is the profile of sigma-algebras provided by the seller to bidders, bidder $i$ believes that bidder $k$ is aware of characteristics in $M^i \cap M^k$. That is, $(\mathcal{I}^k_{M^i \cap M^k})_{k \in N}$ is the profile of sigma-algebras perceived by player $i$, where $\mathcal{I}^k_{M^i \cap M^k}$ is generated by the sets $r^{M^k}_{M^i \cap M^k}(D)$ for $D \in \mathcal{I}^k_{M^k}$. Note that since $1 \in M^i \in \mathcal{M}$ for all $i \in N$, $M^i \cap M^k \neq \emptyset$ for all $i, k \in N$. The assumption that each bidder forms correct beliefs about the awareness of other bidders subject to his own awareness is made to isolate the effect of unawareness from other phenomena. If we were to allow for false beliefs, then it would be unclear whether our results are due to unawareness or false beliefs. That is, this assumption is made to isolate changing awareness from other potentially confounding factors. 

We assume for simplicity that for each bidder $i$, the value of the object is additive in values of characteristics. That is, for the set of characteristics $M' \in \mathcal{M}$, if $(x^i_j)_{j \in M'}$ is the vector of realized values on characteristics $j \in M'$, then $y^i = \sum_{j \in M'} x^i_j$ is bidder $i$'s value of the object. Our approach here is reminiscent of ``pseudotypes'' in scoring auctions; see, for instance, Asker and Cantillon (2008) or Bajari, Houghton, and Tadelis (2014).\footnote{Such an additive specification is plausible for instance in reverse auctions when the value of the object represents the bidder's total cost of producing the object that is additive in the cost of each characteristic.} 

For each bidder $i \in N$ and each subset of characteristics $M^i \in \mathcal{M}$ provided to bidder $i$, define a random variable $Y^i_{M^i}: \Omega \longrightarrow \mathbb{R}$ by if $\omega \in S_{M'}$ then $Y^i_{M^i}(\omega) = \sum_{j \in M^i \cap M'} X^i_j(\omega)$. To see that this definition makes sense, note first that for any $\omega \in S_{M^i}$, $Y^i_{M^i}(\omega) = \sum_{j \in M^i} X^i_j(\omega)$. Second, for any $\omega \in S_{M'}$ with $M^i \subseteq M'$, we also have $Y^i_{M^i}(\omega) = \sum_{j \in M^i} X^i_j(\omega)$. This may be interpreted as follows. Consider a player who is aware of characteristics in $M'$. Such a player knows that bidder $i$ is aware only of characteristics in $M^i \subseteq M'$. Hence, he anticipates that bidder $i$'s value is given by $\sum_{j \in M^i} X^i_j(\omega)$. Third, for any $\omega \in S_{M'}$ with $M^i \supseteq M'$, we have $Y^i_{M^i}(\omega) = \sum_{j \in M'} X^i_j(\omega)$. We can interpret this by considering a player who is aware of characteristics in $M' \subseteq M^i$. Such a player anticipates that bidder $i$ is aware of characteristics in $M'$ but since such a player is unaware of characteristics in $M^i \setminus M'$, he cannot anticipate that bidder $i$ is actually aware of $M^i \supseteq M'$. Hence, he views bidder $i$'s perceived value as given by $\sum_{j \in M'} X^i_j\left((r^{M^i}_{M'})^{-1}(\omega)\right)$ which by the definition of $X^i_j$, $j \in M'$, is $\sum_{j \in M'} X^i_j\left(\omega\right)$. Finally, for any $\omega \in S_{M'}$ such that $M' \nsubseteq M^i$ and $M' \nsupseteq M^i$, $Y^i_{M^i}(\omega) = \sum_{j \in M^i \cap M'} X^i_j(\omega)$. This sum is well-defined because $1 \in M' \cap M^i$. To interpret this consider a player who is aware only of characteristics in $M'$. He must be unaware of characteristics in $M^i \setminus M'$. At the same time, he knows that bidder $i$ is unaware of characteristics in $M' \setminus M^i$. Thus, he envisions bidder $i$'s value to be $\sum_{j \in M^i \cap M'} X^i_j(\omega)$. To sum up, we define the random variable $Y^i_{M^i}$ such that for any player with awareness $M'$, bidder $i$'s total value is consistent with bidder $i$'s awareness and with $M'$.

Next, we define conditional expectations of values. For the profile of sigma-algebras $(\mathcal{I}^i_{M^i})_{i \in N}$ that the seller provides to bidders, an agent with awareness level $M'$ calculates for every bidder $i \in N$ an estimated valuation $(\tilde{Y}^i_{M^i})^{M'}$ for the object based on the profile of sigma-algebras he perceives, $(\mathcal{I}^i_{M^i \cap M'})_{i \in N}$. Let $(\mathcal{I}^i_{M^i \cap M'})^{M'}$ be generated by sets $(r^{M'}_{M^i \cap M'})^{-1}(D)$ for $D \in \mathcal{I}^i_{M^i \cap M'}$. Although an agent with awareness level $M'$ considers $\mathcal{I}^i_{M^i \cap M'}$ to be the information received by bidder $i$, the agent forms beliefs about states in $S_{M'}$. Thus, we need to consider the set of inverse images of elements in $\mathcal{I}^i_{M^i \cap M'}$ in the space $S_{M'}$. This is the set $(\mathcal{I}^i_{M^i \cap M'})^{M'}$. Given $(\mathcal{I}^i_{M^i \cap M'})^{M'}$, the estimated value of bidder $i$ as perceived by an agent with awareness level $M'$ is the random variable $(\tilde{Y}^i_{M^i})^{M'}$ given by the conditional expectation $\mathbb{E}[Y^i_{M^i} \mid (\mathcal{I}^i_{M^i \cap M'})^{M'}]$ defined for all $D \in \left(\mathcal{I}^i_{M^i \cap M'}\right)^{M'}$ by
\begin{eqnarray*} \int_{D} \mathbb{E}[Y^i_{M^i} \mid (\mathcal{I}^i_{M^i \cap M'})^{M'}] (\omega) dP(\omega) & := & \int_{D} \left(\sum_{j \in M^i \cap M'} X_j^i(\omega)\right) d P(\omega) \\ & = & \sum_{j \in M^i \cap M'} \int_{D} X_j^i(\omega) d P(\omega),
\end{eqnarray*} where the last equation follows from the linearity of the expectations operator. 

For a single characteristic $j \in M^i \cap M'$, we write analogously $(\tilde{X}^i_j)^{M'} = \mathbb{E}[X^i_j \mid (\mathcal{I}^i_{M^i \cap M'})^{M'}]$ for bidder $i$'s estimated value of characteristic $j$ as perceived by an agent with awareness level $M'$. By the linearity of conditional expectation, $(\tilde{Y}^i_{M^i})^{M'} = \sum_{j \in M^i \cap M'} (\tilde{X}^i_j)^{M'}$. Whenever no confusion arises, we suppress the superscript $M'$ and write $\tilde{Y}^i_{M^i}$ and $\tilde{X}^i_j$. 

For the random vector of estimated valuations $((\tilde{Y}^1_{M^1})^{M'}, (\tilde{Y}^2_{M^2})^{M'}, ..., (\tilde{Y}^n_{M^n})^{M'})$ on $(S_{M'}, \mathcal{F}_{M'}, P)$ for $M' \in \mathcal{M}$, consider the order statistics from the point of view of an agent with awareness $M'$ and information $((\mathcal{I}^i_{M^i \cap M'})^{M'})_{i \in N}$, $((\tilde{Y}^{(1)})^{M'}, (\tilde{Y}^{(2)})^{M'}, ..., (\tilde{Y}^{(n)})^{M'})$, with $(\tilde{Y}^{(1)})^{M'}$ being the first order statistic given $((\mathcal{I}^i_{M^i \cap M'})^{M'})_{i \in N}$, $(\tilde{Y}^{(2)})^{M'}$ being the second order statistic given $((\mathcal{I}^i_{M^i \cap M'})^{M'})_{i \in N}$ etc.

Note that the order statistics are defined relative to a set of characteristics $M' \in \mathcal{M}$. That is, order statistics may differ by awareness levels.

Given $((\mathcal{I}^i_{M^i \cap M'})^{M'})_{i \in N}$, let $(\mathbf{1}^i)^{M'}$ denote the variable indicating that bidder $i$ wins from an agent's point of view with awareness $M'$ in the weak dominant strategy equilibrium of the second-price auction. This random variable $(\mathbf{1}^i)^{M'}: S_{M'} \longrightarrow \{0, 1\}$ is defined by
\begin{eqnarray*}(\mathbf{1}^i)^{M'}(\omega) & = & \left\{ \begin{array}{cl} 1 & \mbox{ if } (\tilde{Y}^k_{M^k})^{M'}(\omega) \leq (\tilde{Y}^i_{M^i})^{M'}(\omega) \mbox{ for all } k \neq i, \\ 0 & \mbox{ otherwise.} \end{array} \right..
\end{eqnarray*}
If the seller commits to release the information $(\mathcal{I}^1_{M^1}, \mathcal{I}^2_{M^2},..., \mathcal{I}^n_{M^n})$, then bidder $i$'s (ex ante) expected probability of winning from the point of view of an agent with awareness level $M'$ is
\begin{eqnarray*}\mathbb{E}[(\mathbf{1}^i)^{M'}] & = & \int_{S_{M'}} (\mathbf{1}^i)^{M'}(\omega) d P(\omega) \\
& = & \int_{\bigcap_{k \neq i} \left\{\omega \in S_{M'} : (\tilde{Y}^k_{M^k})^{M'}(\omega) \leq (\tilde{Y}^i_{M^i})^{M'}(\omega)\right\}} d P(\omega)\\
& = & P^{M'}\left(\bigcap_{k \neq i} \left\{\omega \in S_{M'} : (\tilde{Y}^k_{M^k})^{M'}(\omega) \leq (\tilde{Y}^i_{M^i})^{M'}(\omega)\right\}\right).
\end{eqnarray*}

Random variable $(\mathbf{1}^i)^{M'}$ is measurable with respect to the sigma-algebra $\bigvee_{i \in N} (\mathcal{I}^i_{M^i \cap M'})^{M'}$.

The first result characterizes the bidders' entry fees that the seller would optimally charge after committing to release information and awareness $(\mathcal{I}^1_{M^1}, \mathcal{I}^2_{M^2},..., \mathcal{I}^n_{M^n})$ as well as the seller's expected revenue from optimal entry fees and the weak dominant strategy equilibrium of the second-price auction. Recall again that when the seller commits to release information $(\mathcal{I}^1_{M^1}, \mathcal{I}^2_{M^2},..., \mathcal{I}^n_{M^n})$, she makes bidder $i$ aware of $M^i$, for $i = 1, ..., n$. Bidder $i$ perceives that the seller committed to release information $(\mathcal{I}^1_{M^1 \cap M^i}, \mathcal{I}^2_{M^2 \cap M^i},..., \mathcal{I}^n_{M^n \cap M^i})$. That is, the seller privately communicates her commitment to information disclosure to each bidder. This communication is truthful w.r.t. the characteristics bidder $i$ is made aware of but it is silent on characteristics of which the seller keeps bidder $i$ unaware. 

\begin{prop}\label{revenue} If the seller commits to release the information $(\mathcal{I}^1_{M^1}, \mathcal{I}^2_{M^2},..., \mathcal{I}^n_{M^n})$ (and thereby raising bidder $i$'s awareness to $M^i$), the highest entry fee that bidder $i \in N$ is willing to pay is
\begin{eqnarray*} e^i = \mathbb{E}\left[\left(Y^i_{M^i} - (\tilde{Y}^{(2)})^{M^i}\right)(\mathbf{1}^i)^{M^i}\right] = \mathbb{E}\left[\left((\tilde{Y}^{(1)})^{M^i} - (\tilde{Y}^{(2)})^{M^i}\right)(\mathbf{1}^i)^{M^i}\right].
\end{eqnarray*}
The expected revenue to the seller is 
\begin{eqnarray*} \mathbb{E}[ (\tilde{Y}^{(1)})^M] + \underbrace{\sum_{i \in N} \left(e^i - (e^i)^M\right)}_{\begin{array}{c} \mbox{Unawareness rents} \\ \mbox{from bidders} \end{array}}
\end{eqnarray*} 
where $(e^i)^M = \mathbb{E}[((\tilde{Y}^{(1)})^M - (\tilde{Y}^{(2)})^M) (\mathbf{1}^i)^{M}]$ is bidder $i$'s entry fee from the point of view of an agent with full awareness $M$. 
\end{prop}

Proposition~\ref{revenue} is inspired by and generalizes Szech (2011). She considers the case in which all bidders are aware of all characteristics. In such a case, the seller's expected revenue is exactly $\mathbb{E}[(\tilde{Y}^{(1)})^M]$. In contrast, we allow the set of characteristics to be different among bidders. This leads to an additional term, $\sum_{i \in N} \left(e^i - (e^i)^M\right)$, in the seller's expected revenue, which is the sum of the differences between a bidder's actual entry fee and the entry fee that the bidder should pay from the point of view of an agent with full awareness $M$. An agent with full awareness $M$ realizes that bidder $i$ with awareness $M^i$ may have a biased perception of his probability of winning and the expected price to be paid upon winning. This creates the difference between $e^i$ and $(e^i)^M$. We may call $e^i - (e^i)^M$ the \emph{unawareness rent} the seller accrues from agent $i$. 

If a bidder is aware of all characteristics in $M$, then the unawareness rent is zero, i.e., $e^i - (e^i)^M = 0$. It follows that the rent $e^i - (e^i)^M$ accrues to the seller only from bidders who are unaware of some characteristic. Yet, asymmetry in awareness is also necessary for accrual of such rents to the seller.

\begin{exa}\label{ex:rent} The following example illustrates the computation of the unawareness rent and the seller's expected revenue. Consider two bidders and two characteristics, $M = \{1, 2\}$. The values of the characteristics $X_1$ and $X_2$ are independently and uniformly distributed on $[0, 5]$ and $[0, 2]$, respectively. We let $f_1$ denote the density of characteristic 1. Applying the convolution theorem, the probability density of $Y = X_1 + X_2$ is 
\begin{eqnarray*} g(y) & = & \left\{ \begin{array}{cl} \frac{1}{10}y & \mbox{ if } 0 \leq y \leq 2, \\ \frac{1}{5} & \mbox{ if } 2 < y \leq 5, \\ \frac{1}{10} (7-y) & \mbox{ if } 5 < y \leq 7. \end{array} \right.
\end{eqnarray*}

We assume that both bidders are aware of the first characteristic. Moreover, each bidder knows that before bidding he will learn his own valuation on the first characteristic but not the valuation of the other bidder. Bidder 1 is also aware of the second characteristic and knows that before bidding he will also learn his valuation for the second characteristic. Bidder 2 remains unaware of the second characteristic and bidder 1 knows that bidder 2 is unaware of the second characteristic. 

Since we have $Y^1 = X_1^1 + X_2^1$ and $Y^2 = X_1^2$, we compute the entry fees that bidders 1 and 2 are willing to pay, respectively, by 
\begin{eqnarray*} e^1 & = & \int^5_0 \int^7_{x^2_1} (y^1 - x^2_1) g(y^1) \, \mathrm{d} y^1 f_1(x^2_1) \, \mathrm{d} x^2_1 =  \frac{109}{75} \\ e^2 & = & \int^5_0 \int^5_{x^1_1} (x^2_1 - x^1_1) f_1(x^2_1) \, \mathrm{d} x^2_1 f_1(x^1_1) \, \mathrm{d} x^1_1  = \frac{5}{6}
\end{eqnarray*} 

Since bidder 1 is aware of both components, there is no unawareness rent from bidder 1. The unawareness rent from bidder 2 is the difference between $e^2$ just computed and $(e^2)^M$. Latter term is
\begin{eqnarray*} (e^2)^M & = & \int^7_0 \int^5_{y^1} (x^2_1 - y^1) f_1(x^2_1) \, \mathrm{d} x^2_1  g(y^1) \, \mathrm{d}y^1 = \frac{34}{75}.
\end{eqnarray*}	Thus, the unawareness rent from bidder 2 is 
\begin{eqnarray*}  e^2 - (e^2)^M & = & \frac{19}{50}.
\end{eqnarray*} The expected revenue to the seller is 
\begin{eqnarray*}
		 & & \mathbb{E}[(\tilde{Y}^{(1)})^M] + e^2 - (e^2)^M \\
		 & = & \int^5_0 \int^7_{x^2_1} y^1 g(y^1) \, \mathrm{d} y^1 f_1(x^2_1) \, \mathrm{d} x^2_1 + \int^7_0 \int^5_{y^1} x^2_1 f_1(x^2_1) \, \mathrm{d} x^2_1  g(y^1) \, \mathrm{d}y^1 + e^2 - (e^2)^M \\
		 & = & \frac{208}{75} + \frac{59}{50} + \frac{5}{6} - \frac{34}{75} \\
		 & = & \frac{416 + 177 + 125 - 68}{150} = \frac{13}{3}
\end{eqnarray*} \hfill $\Box$
\end{exa}

If every bidder is aware of the same set of characteristics, that is $M^i = M' \subseteq M$ for all $i \in N$, the term $\sum_{i \in N} \left(e^i - (e^i)^M\right)$ will drop out because $e^i = (e^i)^M$ for all $i \in N$ (even if $M' \subsetneqq M$). Consequently, the seller's expected revenue is equal to the expected value of the first (highest) order statistics of estimated valuations across spaces. This is essentially Szech (2011)'s result except that in her work it is as if all bidders were aware of the full set of characteristics $M$ to begin with.

\begin{cor}[Szech (2011)]\label{Szech} If the seller commits to release the information $(\mathcal{I}^1_{M'}, \mathcal{I}^2_{M'},..., \mathcal{I}^n_{M'})$ and thereby raising every bidder's awareness to $M' \in \mathcal{M}$, then the seller's expected revenue is $\mathbb{E}[(\tilde{Y}^{(1)})^M]$.
\end{cor}

Note that entry fees are charged before bidders learn their valuations. Under common awareness, the seller therefore extracts the entire expected surplus, and her expected revenue coincides with the expected first order statistic. This differs from the optimal auction of Myerson (1981), where bidders' participation constraints are imposed at the interim stage and the seller leaves information rents to bidders. Proposition~\ref{revenue} shows that under asymmetric awareness the seller may extract more than the expected first order statistic, since unaware bidders misperceive the surplus they are paying for.

\section{Optimally Raising Individual Awareness\label{individual_awareness}}

So far, we characterized optimal entry fees given disclosure. Now we turn to optimal disclosure of individual awareness. Since optimal disclosure of information in auctions has already been studied elsewhere (e.g., Szech, 2011, Eso and Szentes, 2007, and Gershkov, 2009), we focus on optimally raising awareness. W.r.t. information, we assume that the seller has no control over information on the characteristics she may make bidders aware of. That is, when raising awareness of a characteristic, the seller may also provide information on this characteristic. However, while raising awareness of the characteristic is fully under the seller's control, we assume for this case that the seller cannot control the amount of information provided on the characteristics. This is for instance the case when the seller will have some information on some characteristics and raising awareness of a characteristic would also mandate her by law to eventually disclose her information on it. E.g., in real estate transactions in the US, sellers are required by law to disclose information on certain issues. On other issues, the seller is required to disclose all eventually available information \emph{once the buyer asks for it}. Naturally, a buyer can only ask for information when he is aware of it or made aware of it. We will also assume that she will not provide further information on characteristics that bidders are already aware of. 

Throughout this section, we maintain the assumption that each bidder wins the auction with positive probability, that is, $\mathbb{E}[(\mathbf{1}^i)^{M}] > 0$ for every $i \in N$. This rules out cases in which a bidder's estimated valuation is almost surely dominated by some other bidder's estimated valuation. In such cases, raising this bidder's awareness of a characteristic would leave both the expected first order statistic and the unawareness rents unchanged, and all strict inequalities below would have to be replaced by weak ones.

For any bidder $i \in N$ and characteristic $j \in M$, let $\mu_j^i$ denote the expected value of $X_j^i$. Our first observation is that given all the bidders are aware of the same set of characteristics, the seller can improve her expected revenue by making one bidder aware of one more characteristic if the expected value of that characteristic is strictly positive. 

\begin{prop}\label{onemorecharacteristic} Consider any $M' \in \mathcal{M}$ with $M' \subsetneqq M$ and $M^i = M'$ for all $i \in N$, and let $\ell \in M \setminus M'$. If $\mu_{\ell}^1 > 0$, then raising bidder 1's awareness of $\ell$ increases the seller's expected revenue.
\end{prop}

This observation follows from Proposition~\ref{revenue} and two lemmata. When all bidders are exactly aware of $M'$, the seller's expected revenue is $\mathbb{E}[(\tilde{Y}^{(1)})^M]$. Let $\mathbb{E}[(\tilde{Z}^{(1)})^M]$ be the expected first order statistic when bidder $1$ is made aware of characteristic $\ell$. Moreover, let $\sum_{i \in N \setminus \{1\}} \left(e^i_z - (e^i_z)^M\right)$ be the sum of unawareness rents accrued to the seller from all other bidders upon making bidder 1 aware of $\ell$. (Recall that in this case the unawareness rent from bidder 1 is zero.) According to Proposition~\ref{revenue} the seller's expected revenue upon raising bidder $1$'s awareness of $\ell$ is $\mathbb{E}[(\tilde{Z}^{(1)})^M] + \sum_{i \in N \setminus \{1\}} \left(e^i_z - (e^i_z)^M\right)$. We have $\mathbb{E}\left[(\tilde{Z}^{(1)})^M\right] > \mathbb{E}\left[(\tilde{Y}^{(1)})^M\right]$ by the following lemma: 

\begin{lem}\label{inequality} Consider any $M' \in \mathcal{M}$ with $M' \subsetneqq M$ and let $\ell \in M \setminus M'$. If $\mu_{\ell}^1 > 0$, then 
\begin{eqnarray*} \mathbb{E}[\max\{\tilde{Y}^1_{M'} + \tilde{X}^1_{\ell}, \tilde{Y}^2_{M'}, \dots, \tilde{Y}^n_{M'}\}] & > & \mathbb{E}[\max\{\tilde{Y}^1_{M'}, \tilde{Y}^2_{M'}, \dots, \tilde{Y}^n_{M'}\}].
\end{eqnarray*}
\end{lem}

Raising bidder $1$'s awareness of $\ell$ while keeping all other bidders unaware creates unawareness rents from bidders who remain unaware. The next lemma proved in the appendix states that these unawareness rents are positive when bidder $1$'s expected value of characteristic $\ell$ is positive. 

\begin{lem}\label{unawareness_rents} Consider any $M' \in \mathcal{M}$ with $M' \subsetneqq M$ and let $\ell \in M \setminus M'$. If $\mu_{\ell}^1 > 0$, then 
\begin{eqnarray*} \sum_{i \in N \setminus \{1\}} \left(e^i_z - (e^i_z)^M\right) > 0.
\end{eqnarray*} 
\end{lem}

Together, these lemmata prove Proposition~\ref{onemorecharacteristic}. 

What about making further bidders aware of characteristic $\ell$? Raising another bidder's awareness of characteristic $\ell$ involves now a trade-off. Suppose that bidders $1$ to $k$ with $k < n$ are all aware of characteristic $\ell$ already. If the expected value of the characteristic for bidder $k + 1$ is positive, i.e., $\mu_{\ell}^{k + 1} > 0$, then raising awareness of $\ell$ increases the expected first order statistic, which is beneficial to the seller's expected revenue. It also increases the unawareness rents from bidders who remain unaware (if there are still bidders who are unaware) because they now overestimate their probability of winning and underestimate the price they have to pay upon winning even more. Again, this effect is beneficial to the seller. However, the seller also loses the unawareness rent from the bidder whom she made aware, which is the opportunity cost of raising the bidder's awareness of $\ell$. The next observation states this trade-off of raising awareness more formally. 

\begin{prop}\label{onemorebidder} Let $M' \in \mathcal{M}$ with $M' \subsetneqq M$ and let $\ell \in M \setminus M'$. Suppose that bidders $\{1, ..., k\}$ are exactly aware of $M' \cup \{\ell\}$ while bidders $\{k+1, ..., n\}$ are exactly aware of $M'$ only. Raising bidder $k + 1$'s awareness of characteristic $\ell$ increases the seller's expected revenue if $\mu_{\ell}^{k+1} > 0$ and 
\begin{eqnarray} \underbrace{\mathbb{E}\left[(\tilde{Z}^{(1)})^M\right] - \mathbb{E}\left[(\tilde{Y}^{(1)})^M\right]}_{\begin{array}{c} \mbox{Increase in first order} \\ \mbox{statistics}\end{array}} + \underbrace{\sum_{i = k+2}^n \left((e^i)^M - (e^i_z)^M\right)}_{\begin{array}{c}\mbox{Increase in unawareness} \\ \mbox{rents from remaining} \\ \mbox{unaware bidders}\end{array}} & > & \underbrace{e^{k+1} - (e^{k+1})^M}_{\begin{array}{c} \mbox{Loss of unawareness} \\ \mbox{rent from bidder} \\ k + 1 \end{array}} \label{comparison} 
\end{eqnarray} where $(\tilde{Y}^{(1)})^M$ and $(\tilde{Z}^{(1)})^M$ are the first order statistics before and after making bidder $k+1$ aware of $\ell$, respectively, and $e^{i}$ and $e_z^{i}$ are the optimal entry fees of bidder $i$ before and after making bidder $k+1$ aware of $\ell$, respectively. 
\end{prop}

The first term of the l.h.s. of Inequality~(\ref{comparison}) measures the effect of raising one additional bidder's awareness of characteristic $\ell$ on the expected first order statistic. This effect is positive when $\mu_{\ell}^{k+1} > 0$ by the following lemma:  

\begin{lem}\label{lhs1} Let $M' \in \mathcal{M}$ with $M' \subsetneqq M$ and let $\ell \in M \setminus M'$. Suppose that bidders $\{1, ..., k\}$ are exactly aware of $M' \cup \{\ell\}$ while bidders $\{k+1, ..., n\}$ are exactly aware of $M'$ only. If $\mu_{\ell}^{k+1} > 0$, then raising bidder $k + 1$'s awareness of characteristic $\ell$ yields $$\mathbb{E}\left[(\tilde{Z}^{(1)})^M\right] - \mathbb{E}\left[(\tilde{Y}^{(1)})^M\right]  > 0.$$
\end{lem}

The second term of the l.h.s. of Inequality~(\ref{comparison}), $\sum_{i = k+2}^n \left((e^i)^M - (e^i_z)^M\right)$, represents the increase of unawareness rents from remaining unaware bidders when bidder $k + 1$ is made aware of the additional characteristic $\ell$. Recall that unaware bidders ``overpay'' the participation fee because they overestimate their probability of winning and underestimate the amount they would have to pay in case of winning. When bidder $k + 1$ is made aware of the additional characteristic $\ell$ with $\mu_{\ell}^{k + 1} > 0$, then they overestimate their probability of winning even further (because their actual probability of winning decreases) and underestimate the amount they would have to pay in case of winning even further (because the actual expected second order statistic increases). The second term of the l.h.s. of Inequality~(\ref{comparison}) measures the increase of those overpayments, an increase of unawareness rents accrued to the seller from unaware bidders. We prove in the appendix: 

\begin{lem}\label{lhs2} Let $M' \in \mathcal{M}$ with $M' \subsetneqq M$ and let $\ell \in M \setminus M'$. Suppose that bidders $\{1, ..., k\}$ are exactly aware of $M' \cup \{\ell\}$ while bidders $\{k+1, ..., n\}$ are exactly aware of $M'$ only. If $\mu_{\ell}^{k + 1} > 0$, then raising bidder $k + 1$'s awareness of characteristic $\ell$ yields $$\sum_{i = k + 2}^n \left((e^i)^M - (e^i_z)^M\right) > 0.$$
\end{lem}

The r.h.s. of Inequality~(\ref{comparison}), $e^{k+1} - (e^{k+1})^M$, represents the loss of the unawareness rent from bidder $k + 1$ when he is made aware of characteristic $\ell$. As proved in the appendix, the loss is positive when $\mu_{\ell}^{i} > 0$ for $i = 1, ..., k$: 
\begin{lem}\label{rhs1} Let $M' \in \mathcal{M}$ with $M' \subsetneqq M$ and let $\ell \in M \setminus M'$. Suppose that bidders $\{1, ..., k\}$ are exactly aware of $M' \cup \{\ell\}$ while bidders $\{k+1, ..., n\}$ are exactly aware of $M'$ only. If $\mu_{\ell}^i > 0$ for $i = 1, ..., k$, then raising bidder $k + 1$'s awareness of characteristic $\ell$ yields $$e^{k+1} - (e^{k+1})^M  > 0.$$
\end{lem}

One may wonder whether $\mu_{\ell}^i > 0$ is necessary for the above observations to hold. In particular, for a seller to improve her revenue by making bidders aware of a characteristic without control of information, is it necessary that this characteristic has a positive expected value? The following examples provide negative answers.

\begin{exa}\label{ex:negmean} There are cases in which the seller optimally raises a bidder's awareness of characteristic $\ell$ even though $\mu_\ell < 0$. Suppose that for both bidders, the random variable for characteristic 1, $X_1$, is uniformly distributed on $[0, 5]$, and for characteristics 2, $X_2$, is uniformly distributed on $[-6, 5]$ with $\mu_2 = -\frac{1}{2}$. Then, $\mathbb{E}[\max\{X^1_1, X^2_1\}] = \frac{10}{3}$. The distribution of $Y = X_1 + X_2$ is
\begin{eqnarray*} g(y) & = & \left\{ \begin{array}{cl} \frac{1}{55}(6+y) & \mbox{ if } -6 \leq y \leq -1, \\ \frac{1}{11} & \mbox{ if } -1 < y \leq 5, \\ \frac{1}{55} (10-y) & \mbox{ if } 5 < y \leq 10. \end{array} \right..
\end{eqnarray*} Suppose the seller raises bidder 1's awareness of the second characteristic and provides full information about it. Then we have
\begin{eqnarray*}
& & \mathbb{E}\left[\max\{X^1_1+X^1_2, X^2_1\}\right] \\
& = & \int^5_0 \int^{10}_{x_1} y g(y) \, \mathrm{d}y f(x_1) \, \mathrm{d}x_1 + \int^0_{-6} \int^5_0 x_1 f(x_1) \, \mathrm{d}x_1 g(y) \, \mathrm{d}y + \int^5_0 \int^5_y x_1 f(x_1) \, \mathrm{d}x_1 g(y) \, \mathrm{d}y \\
& = & \frac{505}{132} \approx 3.826,
\end{eqnarray*} which is greater than $\mathbb{E}[\max\{X^1_1, X^2_1\}]$. This example just shows that there are cases when the expected value of the second characteristic is negative, we can still have $\mathbb{E}[\max\{X^1_1 + X^1_2, X^2_1\}] > \mathbb{E}[\max\{X^1_1, X^2_1\}]$. \hfill $\Box$
\end{exa}

The next example shows that Example~\ref{ex:negmean} is not pathological. We obtain similar examples in the ``normal'' case.

\begin{exa}\label{ex:normal} Suppose that for both bidders, the random variables for characteristics 1 and 2, $X_1$ and $X_2$, are independent normal random variables distributed as $N(\mu_1, \sigma^2_1)$ and $N(\mu_2, \sigma^2_2)$. Then $X_1 + X_2$ is distributed as $N(\mu_1 + \mu_2, \sigma^2_1 + \sigma^2_2)$. Applying the results from Clark (1961), we have the expected values for the two highest order statistics as follows:
\begin{eqnarray*}
\mathbb{E}[\max\{X^1_1, X^2_1\}] = \mu_1 + \frac{\sigma_1}{\sqrt{\pi}}
\end{eqnarray*}
\begin{eqnarray*}
\mathbb{E}[\max\{X^1_1 + X^1_2, X^2_1\}] = \mu_1 + \mu_2 \Phi \left(\frac{\mu_2}{\sqrt{2\sigma^2_1 + \sigma^2_2}}\right) + \sqrt{2\sigma^2_1 + \sigma^2_2} \phi \left(\frac{\mu_2}{\sqrt{2\sigma^2_1 + \sigma^2_2}}\right),
\end{eqnarray*}
where $\Phi$ and $\phi$ are the CDF and PDF of a standard normal distribution. The difference between $\mathbb{E}[\max\{X^1_1 + X^1_2, X^2_1\}]$ and $\mathbb{E}[\max\{X^1_1, X^2_1\}]$, which is equal to $\mu_2 \Phi \left(\frac{\mu_2}{\sqrt{2\sigma^2_1 + \sigma^2_2}}\right) + \sqrt{2\sigma^2_1 + \sigma^2_2} \phi \left(\frac{\mu_2}{\sqrt{2\sigma^2_1 + \sigma^2_2}}\right)-\frac{\sigma_1}{\sqrt{\pi}}$, does not depend on $\mu_1$, but it increases in $\mu_2$ and $\sigma_2$, and decreases in $\sigma_1$.\footnote{Let $D$ be the difference. We have $\frac{\partial D}{\partial \mu_2} = \Phi \left(\frac{\mu_2}{\sqrt{2\sigma^2_1 + \sigma^2_2}}\right) > 0$, $\frac{\partial D}{\partial \sigma_2} = \frac{\sigma_2}{\sqrt{2\sigma^2_1 + \sigma^2_2}} \phi \left(\frac{\mu_2}{\sqrt{2\sigma^2_1 + \sigma^2_2}}\right) > 0$, and $\frac{\partial D}{\partial \sigma_1} = \frac{2\sigma_1}{\sqrt{2\sigma^2_1 + \sigma^2_2}} \phi \left(\frac{\mu_2}{\sqrt{2\sigma^2_1 + \sigma^2_2}}\right) - \frac{1}{\sqrt{\pi}} < 0$ since $\frac{2\sigma_1}{\sqrt{2\sigma^2_1 + \sigma^2_2}} < \sqrt2$ and $\phi \left(\frac{\mu_2}{\sqrt{2\sigma^2_1 + \sigma^2_2}}\right) < \frac{1}{\sqrt{2\pi}}$.} When $\mu_2 = 0$, the difference becomes
\begin{eqnarray*}
\frac{\sqrt{2\sigma^2_1 + \sigma^2_2}}{\sqrt{2\pi}} - \frac{\sigma_1}{\sqrt{\pi}} \geq 0
\end{eqnarray*} with equality when $\sigma_2 = 0$, which would make $X_2$ a constant. (This case is ruled out almost surely by assumption). By monotonicity, if $\mu_2 > 0$,
\begin{eqnarray*}
& & \mathbb{E}[\max\{X^1_1 + X^1_2, X^2_1\}] - \mathbb{E}[\max\{X^1_1, X^2_1\}] \\
& = & \mu_2 \Phi \left(\frac{\mu_2}{\sqrt{2\sigma^2_1 + \sigma^2_2}}\right) + \sqrt{2\sigma^2_1 + \sigma^2_2} \phi \left(\frac{\mu_2}{\sqrt{2\sigma^2_1 + \sigma^2_2}}\right) - \frac{\sigma_1}{\sqrt{\pi}} \\
& > & \frac{\sqrt{2\sigma^2_1 + \sigma^2_2}}{\sqrt{2\pi}} - \frac{\sigma_1}{\sqrt{\pi}}.
\end{eqnarray*}

In a two-bidder setting, $\mu_2 > 0$ is a sufficient condition for the inequality. The necessary condition also depends on $\sigma_1$ and $\sigma_2$. In the extreme case when $\sigma_1 = 0$ and $\sigma_2 = 0$ (and thus information about those random variables would not play any role), $\mu_2 > 0$ is the sufficient and necessary condition. (Again, we ruled out earlier by assumption that random variables are almost surely constant.) \hfill $\Box$
\end{exa}

\section{Optimally Raising Common Awareness\label{info}} 

So far, we considered a seller who can make bidders aware individually via targeted disclosure of awareness to a particular bidder. Now we consider the case where the seller is constrained to disclose awareness to \emph{all} bidders if any. In such a case, awareness is never asymmetric, and unawareness rents are therefore always zero by Corollary~\ref{Szech}. We focus on two sub-cases: First, we consider the case of raising common awareness of a characteristic without providing any information on it. This is followed by a second case in which the seller raises common awareness of a characteristic and provides full information about the characteristic. We assume for this section that the valuations for each characteristic are not only independently distributed across bidders but also identically distributed across bidders. For any characteristic $\ell \in M$, let $\mu_{\ell}$ denote the (common) expected value of $X_{\ell}$.

\subsection{No Information\label{no_info}} 

Raising awareness of a characteristic without providing any information on the characteristic is relevant, for instance, when bidders bid for access to a user and the seller is forbidden by privacy laws to provide any information about the user. In such a case, the seller could at least raise the bidders' awareness of all characteristics that she collects information on without disclosing any of this information. Subsequently, bidders may now consider the expected value of some characteristics that they did not take into account previously. Another instance would be a procurement setting in which the procurement officer raises awareness of potential events that might affect the bidders' costs without having any information on these events. Yet, another example would be the disclosure of risk factors that might affect the future profitability of a take over candidate in financial markets. Typically, such risk factors are just listed in official filings without information as to their likelihood. 

\begin{prop}\label{privacy} Suppose the seller can only commit to disclosure of information $(\mathcal{I}^1_{M'}, \mathcal{I}^2_{M'},..., \mathcal{I}^n_{M'})$ that is completely uninformative about characteristics $M'$ but allows for raising common awareness of any subsets of characteristics $M' \in \mathcal{M}$. For any characteristic $\ell \in M$, the optimal common awareness raising strategy of the seller raises all bidders' awareness of characteristic $\ell$ if and only if $\mu_\ell > 0$. 
\end{prop}

\subsection{Full Information\label{full_info}} 

Raising common awareness under full information is relevant in cases where the seller will eventually have full information and is required by law to disclose all information. Bidders can only ask for that information (after paying their entry fees) when they are aware of it or made aware of it. The next proposition gives a sufficient and a necessary condition for keeping bidders unaware under full information. 

\begin{prop}\label{unaware} Suppose the seller can commit to any disclosure of information $(\mathcal{I}^1_{M'}, \mathcal{I}^2_{M'},..., \mathcal{I}^n_{M'})$ with common awareness of characteristics in $M'$ for any $M' \in \mathcal{M}$ but full information about characteristics disclosed. For any $\ell \in M$, the optimal awareness raising strategy of the seller keeps all bidders unaware of $\ell$ if $\mathbb{E}[\max \{X^1_{\ell}, X^2_{\ell}, \dots, X^n_{\ell}\}] < 0,$ and only if $\mu_{\ell} < 0$.
\end{prop}

Proposition~\ref{unaware} is not a characterization. The sufficient and the necessary condition differ whenever $\mu_{\ell} < 0 \leq \mathbb{E}[\max\{X^1_{\ell}, \dots, X^n_{\ell}\}]$, and in this region the seller's decision depends on the distribution of the characteristics in $M'$ rather than on the distribution of $X_{\ell}$ alone. This is the case despite our assumption that the values of all characteristics are drawn independently across characteristics and bidders. Independence renders the valuations additively separable across characteristics, but the seller's expected revenue is the expectation of the first order statistic, and the maximum is not additive. Which bidder's realization of $X_{\ell}$ enters the first order statistic is determined by the realizations of the characteristics in $M'$, so the two cannot be evaluated in isolation. In contrast, under no information every bidder's estimated valuation shifts by the same constant $\mu_{\ell}$, the identity of the winner is unaffected, and Proposition~\ref{privacy} is therefore a characterization in $\mu_{\ell}$. The following example illustrates this.

\begin{exa}\label{ex:notchar} Consider two bidders, $M' = \{1\}$, and a further characteristic $\ell$. Let $X^i_1$ be uniformly distributed on $[0, 100]$ and let $X^i_{\ell}$ take the value $10$ with probability $\frac{1}{10}$ and the value $-2$ with probability $\frac{9}{10}$, independently across bidders and independent of $X^i_1$. Then $\mu_{\ell} = -\frac{4}{5}$ while
\begin{eqnarray*} \mathbb{E}[\max\{X^1_{\ell}, X^2_{\ell}\}] & = & 10\left(1 - \left(\tfrac{9}{10}\right)^2\right) - 2 \left(\tfrac{9}{10}\right)^2 \; = \; \tfrac{7}{25} \; > \; 0,
\end{eqnarray*} so the sufficient condition of Proposition~\ref{unaware} fails. Nevertheless, raising common awareness of $\ell$ strictly decreases the seller's expected revenue:
\begin{eqnarray*} \mathbb{E}\left[\max\{X^1_1 + X^1_{\ell}, X^2_1 + X^2_{\ell}\}\right] - \mathbb{E}\left[\max\{X^1_1, X^2_1\}\right] & = & -\tfrac{10556}{15625} \; \approx \; -0.676 \; < \; 0,
\end{eqnarray*} where $\mathbb{E}\left[\max\{X^1_1 + X^1_{\ell}, X^2_1 + X^2_{\ell}\}\right] = \frac{3093332}{46875} \approx 65.991$ and $\mathbb{E}[\max\{X^1_1, X^2_1\}] = \frac{200}{3} \approx 66.667$. Hence the seller optimally keeps both bidders unaware of $\ell$ even though $\mathbb{E}[\max\{X^1_{\ell}, X^2_{\ell}\}] > 0$. The reason is that $X_1$ is dispersed relative to $X_{\ell}$, so the identity of the winner is almost entirely determined by $X_1$ and hence is essentially uninformative about the realization of $X_{\ell}$ of the winning bidder. The value of $X_{\ell}$ that is added to the first order statistic is then close to a draw of $X_{\ell}$ selected independently of its realization rather than to the maximum of the two draws, and the gain from raising awareness is close to $\mu_{\ell}$ rather than to $\mathbb{E}[\max\{X^1_{\ell}, X^2_{\ell}\}]$. Note that the example is not sensitive to the scale of $X_1$: increasing the upper bound of the support of $X_1$ drives the difference further towards $\mu_{\ell} = -\frac{4}{5}$. \hfill $\Box$
\end{exa}

\section{Discussion\label{discussion}}

\subsection{Winner's Curse Due to Unawareness\label{winners_curse}}

If the seller optimally raises awareness of characteristics, then any characteristic of which she keeps bidders unaware has a negative expected value. This will lead to an effect akin to the winner's curse. This winner's curse is not due to common values but due to unawareness of characteristics with negative expected value and occurs even in our setting with independent valuations. 

To illustrate, suppose that all bidders are aware of characteristics in the set $M' \in \mathcal{M}$ and there is one characteristic $\ell \in M \setminus M'$ represented by a random variable $X_{\ell}$ that is independently and identically distributed among all bidders. Assume that $\mu_\ell < 0$.  If the seller keeps bidders unaware of $\ell$, her expected revenue will be $\mathbb{E}[Y^{(1)}]$ such that for all $i \in N$, $Y^i = \sum_{j \in M'} X^i_j$ is based on the same set of characteristics $M'$ that does not include $\ell$. If bidder $i$ were aware of $\ell$, his valuation of the object would be $Y^i_{M'} + X^i_{\ell}$. In such a case, bidder $i$'s actual expected payoff from the auction would be
\begin{eqnarray*} \mathbb{E}\left[\left(Y^i_{M'} + X^i_{\ell} - Y^{(2)}\right)\mathbf{1}^i\right] - e^i & = & \mathbb{E}\left[\left(Y^{(1)} - Y^{(2)}\right)\mathbf{1}^i\right] - e^i + \mathbb{E}[X^i_{\ell}] \, \mathbb{E}[\mathbf{1}^i],
\end{eqnarray*} where the last equality uses the independence of $X^i_{\ell}$ from the characteristics in $M'$ and hence from the event that bidder $i$ wins. Since $e^i = \mathbb{E}[(Y^{(1)} - Y^{(2)})\mathbf{1}^i]$ by Proposition~\ref{revenue}, bidder $i$'s perceived expected payoff is zero, while his actual expected payoff is $\mu_{\ell} \, \mathbb{E}[\mathbf{1}^i] < 0$. This is a form of winner's curse due to unawareness of characteristics that negatively affect bidders. 

Bidders may realize that the seller has no incentive to raise awareness of characteristics that negatively affect bidders. That is, they may become aware that there may or may not be characteristics of which they are unaware. Further, they may realize that if there is such a characteristic, then it has a negative expected value. Each bidder may form different beliefs about what he might be unaware of and adjust his bids accordingly. At the end the most optimistic among the bidders might win the auction and may still suffer a winner's curse. 

We emphasize that the winner's curse in our setting is different from the standard setting. The winner's curse in standard common value auctions occurs because the winner is the interim most optimistic bidder even with bidders that are unbiased ex ante. Our case features independent values. Bidders are also unbiased w.r.t. characteristics they are aware of but because the auctioneer keeps bidders unaware of characteristics with negative expected value, they essentially misperceive their expected value of the object.

\subsection{How is Awareness Different from Information?\label{awareness_vs_information}} 

Conceptually, awareness is very different from information. Clearly, there is a difference between not being able to think of a random variable $X_{\ell}$ (aka unawareness of $\ell$) and realizing that there is a random variable $X_{\ell}$ but having no information about it and thus only taking into account $\mu_{\ell}$ (aka lack of information about $X_{\ell}$). 

In the context of second-price auctions with costless information, disclosure of information is very different from disclosure of awareness. We have seen that disclosure of awareness involves a delicate trade-off between the effect on the expected first order statistic and the effect on unawareness rents. Optimal disclosure of costless information on the other hand leads to full disclosure. In our context, given common awareness of a set of characteristics, full information disclosure is optimal for the seller. 

\begin{prop}\label{perfectinfo} Given a set of characteristics $M' \in \mathcal{M}$, suppose the seller can commit to any disclosure of information $(\mathcal{I}^1_{M'}, \mathcal{I}^2_{M'},..., \mathcal{I}^n_{M'})$ (i.e., any information structure for which all bidders' awareness levels are raised to a given common set of characteristics $M' \in \mathcal{M}$). Then committing to full information on all characteristics in $M'$ maximizes the seller's expected revenue among all $(\mathcal{I}^1_{M'}, \mathcal{I}^2_{M'},..., \mathcal{I}^n_{M'})$.
\end{prop}

While the result generalizes an observation of Szech (2011) for situations of common full awareness, the arguments in the appendix are the same as in Szech (2011). Similar results and arguments have also been presented by Eso and Szentes (2007) and Gershkov (2009). 

The comparison of Propositions~\ref{privacy} and~\ref{unaware} points to a further difference between awareness and information. As noted in Section~\ref{full_info}, raising common awareness of a characteristic without providing information on it shifts every bidder's estimated valuation by the same constant and leaves the identity of the winner unaffected, so that raising awareness is profitable if and only if $\mu_{\ell} > 0$. Once awareness is accompanied by information, bidders learn their individual realizations of $X_{\ell}$, and these realizations also determine who wins. Whether raising awareness pays then depends on the joint configuration of $X_{\ell}$ and the characteristics of which bidders are already aware, and not on the distribution of $X_{\ell}$ alone; see Example~\ref{ex:notchar}. Raising awareness without information shifts the location of the bidders' valuations, whereas raising awareness with information also changes the selection among them.

More generally, raising awareness differs from disclosure of verifiable information (e.g., Milgrom, 1981 and Milgrom and Roberts, 1986; see Milgrom 2008 for a survey). Raising awareness is more an ``eye opener'' about the general existence of a dimension previously not considered. Disclosure of verifiable information is about ruling out possible values along a preconceived dimension. For instance, information unraveling in the literature on disclosure of verifiable information relies crucially on the receiver's ability to reason about which information she did \emph{not} receive. Such a reasoning requires being aware of the information she did not receive. Heifetz, Meier, and Schipper (2021) show that if the receiver is unaware of a dimension of quality, then she cannot infer a low quality from non-disclosure by the sender and hence unraveling may fail under unawareness. Li and Schipper (2025) test this theory and find support for it. However, raising awareness is similar to disclosure of verifiable information in that it cannot be neglected by the receiver. Once aware of a dimension, the receiver must form beliefs about the values along the dimension.

\subsection{Related Literature on Unawareness\label{unaw_lit}} 

Besides the literature on information disclosure in auctions, we contribute to the study of unawareness in contracting. We make use of unawareness type spaces, albeit in a very restricted form, by Heifetz, Meier, and Schipper (2013a), Bayesian games with unawareness by Meier and Schipper (2014), and dynamic games with unawareness by Heifetz, Meier, and Schipper (2013b). Francetich and Schipper (2026) study a screening problem in which the agent is more aware than the principal and has to decide whether or not to raise the principal's awareness before the principal offers a menu of screening contracts. They show that when the principal is unaware of high marginal cost types only, then the agent happily raises her awareness while this is not the case when the principal is unaware of low marginal cost types only. Herweg and Schmidt (2020) study a procurement problem in which the seller may be aware of some ex post verifiable design flaws. They propose an efficient two-stage mechanism with a neutral arbitrator that does not require an ex ante description of design flaws. Grant, Kline, and Quiggin (2012) discuss disagreements arising from asymmetric awareness in contracting. Pram and Schipper (2026) study general efficient mechanism design under unawareness for quasi-linear environments. They devise dynamic elaboration direct mechanisms for efficient implementation with no-deficit. von Thadden and Zhao (2012) study a principal-agent moral hazard problem in which the principal is aware of actions of which the agent is unaware. When contemplating whether or not to make the agent aware, the principal faces a trade-off between getting a better action and saving on information rents due to additional incentive compatibility constraints. Auster (2013) studies a principal-agent moral hazard problem in which the principal is aware of contingencies of which the agent is unaware but whose realization is informative about the agent's actions. In the optimal contract, the principal faces a trade-off between exploiting the agent's unawareness and using said contingencies in order to provide incentives. Filiz-Ozbay (2012) studies a risk neutral insurer who is aware of some contingencies that the insuree is unaware. The insurer has an incentive to mention only some contingencies in a contract while remaining silent on others. Auster and Pavoni (2024) and Lei and Zhao (2021) feature an agent with higher awareness level than the principal but in the context of optimal delegation. In Auster and Pavoni (2024), the agent is aware of both the set of his actions and their performance, and only reveals extreme actions. In Lei and Zhao (2021), the agent only partially reveals contingencies of which the principal is unaware. Principals who are unaware of more contingencies delegate a large set of projects. Finally, in the field of value-based business strategy, Bryan, Ryall, and Schipper (2022) devise cooperative games with incomplete information and unawareness for studying surplus creation and surplus appropriation within contracting relationships.

\appendix
\renewcommand{\thesubsection}{A.\arabic{subsection}}
\setcounter{section}{0}

\section{Additional Mathematical Results for the Model\label{additional}}

We state two preliminary lemmata that serve as basis for our analysis. 

\begin{lem}\label{measurableX} For every $M', M'' \in \mathcal{M}$ with $M' \subseteq M''$, $X_{M'}$ is measurable on $S_{M''}$.
\end{lem}

\noindent \textsc{Proof. } Since $X_{M'}$ is a measurable random variable on $S_{M'}$, we have for every $x_{M'} \in \mathbb{R}^{n \cdot |M'|}$ that $\left\{\omega \in S_{M'} : X_{M'}(\omega) \leq x_{M'} \right\} \in \mathcal{F}_{M'}$. From the extension of $X_{M'}$ to $S_{M''}$ follows that $\left\{\omega \in S_{M'} : X_{M'}\left((r^{M''}_{M'})^{-1}(\omega)\right) \leq x_{M'}\right\} \in \mathcal{F}_{M'}$. Since projections are measurable, we have $(r^{M''}_{M'})^{-1}\left(\left\{\omega \in S_{M'} : X_{M'}\left((r^{M''}_{M'})^{-1}(\omega)\right) \leq x_{M'}\right\}\right) \in \mathcal{F}_{M''}$, which is equivalent to $\left\{\omega \in S_{M''} : X_{M'}\left(\omega\right) \leq x_{M'}\right\} \in \mathcal{F}_{M''}$. \hfill $\Box$\\

The following lemma shows that lattice of measurable spaces with the projective system of probability measures captures the intuition analogous to a standard probability space.

\begin{lem}\label{P} For any $M', M'', M''' \in \mathcal{M}$ with $M' \subseteq M'' \cap M'''$ and any $x_{M'} \in \mathbb{R}^{n \cdot |M'|}$,
\begin{eqnarray*} P^{M''} (\{\omega \in \Omega : X_{M'}(\omega) \leq x_{M'}\}) = P^{M'''}(\{\omega \in \Omega: X_{M'}(\omega) \leq x_{M'}\}).
\end{eqnarray*}
\end{lem}

\noindent \textsc{Proof. } For any $M', M'', M''' \in \mathcal{M}$ with $M' \subseteq M'' \cap M'''$ and any $x_{M'} \in \mathbb{R}^{n \cdot |M'|}$, we have
\begin{eqnarray} \lefteqn{P^{M'' \cap M'''}(\{\omega \in \Omega: X_{M'}(\omega) \leq x_{M'}\}) } \nonumber \\  & = & P^{M''}_{|M'' \cap M'''}(\{\omega \in \Omega: X_{M'}(\omega) \leq x_{M'}\}) \label{P1} \\
	& = & P^{M''}\left((r^{M''}_{M'' \cap M'''})^{-1}\left(\left\{\omega \in S_{M'} : X_{M'}(\omega)\leq x_{M'}\right\} \right)\right) \label{P2} \\
	& = & P^{M''}\left(\left\{\omega \in \Omega : X_{M'}(\omega) \leq x_{M'}\right\} \cap S_{M''} \right) \label{P3} \\
	& = & P^{M''}\left(\left\{\omega \in \Omega : X_{M'}(\omega) \leq x_{M'}\right\} \right), \nonumber
\end{eqnarray} where equation~(\ref{P1}) follows from the fact that $P$ is a projective system of probability measures, equation~(\ref{P2}) follows from the definition of marginal, and equation~(\ref{P3}) follows from the extension of the random variable $X_{M'}$ to $S_{M''}$.  By analogous arguments, we have $P^{M'' \cap M'''}(\{\omega \in \Omega: X_{M'}(\omega) \leq x_{M'}\}) = P^{M'''}\left(\left\{\omega \in \Omega : X_{M'}(\omega) \leq x_{M'}\right\} \right)$. Hence $P^{M''}(\{\omega \in \Omega: X_{M'}(\omega) \leq x_{M'}\}) = P^{M'''}\left(\left\{\omega \in \Omega : X_{M'}(\omega) \leq x_{M'}\right\} \right)$. \hfill $\Box$\\ 

Next, we collect some mathematical results that are useful for applying our model using specific distributions. 

Denote by $(G^i_{M^i})^{M'}$ the distribution of $Y^i_{M^i}$ from the point of view of a player who is aware of $M'$. That is,
\begin{eqnarray*} (G^i_{M^i})^{M'}(y^i) = P^{M'}\left(\left\{\omega \in S_{M'} : Y_{M^i}^i(\omega) \leq y^i\right\}\right).
\end{eqnarray*}
We like to obtain an expression for $(G^i_{M^i})^{M'}$ in terms of the distributions of $X_j^i$'s. To this extent, define an isotone bijection $b^{M^i \cap M'} : M^i \cap M' \longrightarrow \{1, ..., |M^i \cap M'|\}$ by $b^{M^i \cap M'}(1) = 1$, and $b^{M^i \cap M'}(j) > b^{M^i \cap M'}(k)$ if and only if $j > k$. This bijection just renumbers characteristics in $M^i \cap M'$ consecutively. Using this renumbering and the assumption of independence of random variables across characteristics, the following lemma follows directly from the convolution theorem (see for instance, Chung, 2001, Theorem 6.1.1., Corollary p. 153, and Theorem 6.1.4.):
\begin{lem}\label{convolution} Writing $b = b^{M^i \cap M'}$ and $\iota = |M^i \cap M'|$,
\begin{eqnarray*} (G^i_{M^i})^{M'}(y^i) & = & \int \cdots \int F^i_{b^{-1}(1)}\left(y^i - x^i_2 - \ldots - x^i_{\iota}\right) d F^i_{b^{-1}(2)}\left(x^i_2\right) \cdots d F^i_{b^{-1}(\iota)}\left(x^i_{\iota}\right)
\end{eqnarray*}
\end{lem}

Denote the distribution of $\tilde{Y}^i_{M^i}$ from the point of view of a player with awareness $M'$ given $((\mathcal{I}^i_{M^i \cap M'})^{M'})_{i \in N}$ by $(\tilde{G}^i_{M^i})^{M'}$ such that
\begin{eqnarray*}(\tilde{G}^i_{M^i})^{M'}(y^i) = P^{M'}(\{\omega \in S_{M'}: (\tilde{Y}^i_{M^i})^{M'}(\omega) \leq y^i \}).
\end{eqnarray*} Denote the probability distributions of $(\tilde{Y}^{(1)})^{M'}$ and $(\tilde{Y}^{(2)})^{M'}$ on $S_{M'}$ given $((\mathcal{I}^i_{M^i \cap M'})^{M'})_{i \in N}$ by $(\tilde{G}^{(1)})^{M'}$ and $(\tilde{G}^{(2)})^{M'}$, respectively. The characterizations of these distributions follow from G\"{u}ng\"{o}r, Bulut, and \c{C}al\i k (2009, Result 2.4):

\begin{lem}\label{order_statistics}
\begin{eqnarray*} (\tilde{G}^{(1)})^{M'}(y) & = & \prod_{i \in N} (\tilde{G}^i_{M^i})^{M'} (y) \\
(\tilde{G}^{(2)})^{M'}(y) & = & \prod_{i \in N} (\tilde{G}^i_{M^i})^{M'}(y) + \sum_{i \in N} \left[ (1 - (\tilde{G}^i_{M^i})^{M'}(y)) \prod_{k \in N \setminus \{i\}}(\tilde{G}^k_{M^k})^{M'}(y)\right]
\end{eqnarray*}
\end{lem}

\noindent \textsc{Proof. } Recall that the permanent of a $n \times n$ square matrix $\bold{A} = (a_{ij})$ is defined by
$$per \bold A = \sum_{\sigma \in P_n} \prod_{i=1}^n a_{i \sigma(i)},$$
where $P_n$ is the set of permutations of $\{1, 2, \dots, n\}$. Then the permanent is equivalent to the determinant except that all signs in the expansion are positive. From Result 2.4 of G\"{u}ng\"{o}r, Bulut, and \c{C}al\i k (2009) for the distribution function of order statistics with ascending values, we have the distribution function for the $(n+1-r)$-th order statistic for descending values
\begin{eqnarray*} (\tilde{G}^{(n+1-r)})^{M'}(y) & = & \Pr[(\tilde{Y}^{(n+1-r)})^{M'} \leq y] \\
	& = & \sum^n_{m = r} \frac{1}{m!(n-m)!} per \left[\underbrace{\bold{(\tilde{G})}^{M'}(y)}_\text{m} \; \underbrace{\bold{1 - (\tilde{G})}^{M'}(y)}_\text{n-m}\right]
\end{eqnarray*}
where $\bold{(\tilde{G})}^{M'}(y) := \left((\tilde{G}^1_{M^1})^{M'}(y), (\tilde{G}^2_{M^2})^{M'}(y), \dots, (\tilde{G}^n_{M^n})^{M'}(y)\right)'$ and $\bold{1 - (\tilde{G})}^{M'}(y) := \\ \left(1-(\tilde{G}^1_{M^1})^{M'}(y), 1-(\tilde{G}^2_{M^2})^{M'}(y), \dots, 1-(\tilde{G}^n_{M^n})^{M'}(y)\right)'$ are column vectors, and there are $m$ columns of $\bold{(\tilde{G})}^{M'}(y)$ and $n-m$ columns of $\bold{1 - (\tilde{G})}^{M'}(y)$ in the $permanent$ operator. Therefore, when $r = n$ we have the distribution function for the highest order statistic
\begin{eqnarray*} (\tilde{G}^{(1)})^{M'}(y) & = & \Pr[(\tilde{Y}^{(1)})^{M'} \leq y] \\
	& = & \frac{1}{n!0!} per \left[\underbrace{\bold{(\tilde{G})}^{M'}(y)}_\text{n}\right] \\
	& = & \prod_{i \in N} (\tilde{G}^i_{M^i})^{M'} (y),
\end{eqnarray*}
and when $r = n-1$ we have the distribution function for the second-highest order statistic
\begin{eqnarray*} (\tilde{G}^{(2)})^{M'}(y) & = & \Pr[(\tilde{Y}^{(2)})^{M'} \leq y] \\
	& = & \sum^n_{m = n-1} \frac{1}{m!(n-m)!} per \left[\underbrace{\bold{(\tilde{G})}^{M'}(y)}_\text{m} \; \underbrace{\bold{1 - (\tilde{G})}^{M'}(y)}_\text{n-m}\right] \\
	& = & \frac{1}{(n-1)!1!} per \left[\underbrace{\bold{(\tilde{G})}^{M'}(y)}_\text{n-1} \; \underbrace{\bold{1 - (\tilde{G})}^{M'}(y)}_\text{1}\right] + \frac{1}{n!0!} per \left[\underbrace{\bold{(\tilde{G})}^{M'}(y)}_\text{n}\right] \\
	& = & \sum_{i \in N} \left[ (1 - (\tilde{G}^i_{M^i})^{M'}(y)) \prod_{k \in N \setminus \{i\}}(\tilde{G}^k_{M^k})^{M'}(y)\right] + \prod_{i \in N} (\tilde{G}^i_{M^i})^{M'}(y).
\end{eqnarray*} \hfill $\Box$

\section{Proofs\label{proofs}}

\subsection*{Proof of Proposition~\ref{revenue}}

The entry fee for bidder $i$ is his expected payoff from the weak dominant strategy equilibrium of the second-price auction with bidders informed by $((\mathcal{I}^k_{M^k \cap M^i})^{M^i})_{k \in N}$. We want to show that it is exactly equal to $e^i = \mathbb{E}\left[\left((\tilde{Y}^{(1)})^{M^i} - (\tilde{Y}^{(2)})^{M^i}\right)(\mathbf{1}^i)^{M^i}\right]$. We have
\begin{eqnarray*}
	\mathbb{E}\left[(\tilde{Y}^{(1)})^{M^i} (\mathbf{1}^i)^{M^i}\right] & = & \mathbb{E}\left[(\tilde{Y}^i_{M^i})^{M^i} (\mathbf{1}^i)^{M^i}\right] \\
	& = & \mathbb{E}\left[\mathbb{E}[Y^i_{M^i} \mid (\mathcal{I}^i_{M^i})^{M^i}] (\mathbf{1}^i)^{M^i}\right] \\
	& = & \mathbb{E}\left[\mathbb{E}[Y^i_{M^i} \mid \bigvee_{k \in N}(\mathcal{I}^k_{M^k \cap M^i})^{M^i}] (\mathbf{1}^i)^{M^i}\right] \\
	& = & \mathbb{E}\left[\mathbb{E}[Y^i_{M^i} (\mathbf{1}^i)^{M^i} \mid \bigvee_{k \in N}(\mathcal{I}^k_{M^k \cap M^i})^{M^i}]\right] \\
	& = & \mathbb{E}\left[Y^i_{M^i} (\mathbf{1}^i)^{M^i}\right],
\end{eqnarray*} where the first equality follows because bidder $i$'s estimated valuation is the highest if he wins the auction, the second equality is by the definition of $(\tilde{Y}^i_{M^i})^{M^i}$, the third equality is due to independence, the fourth equality follows from $(\mathbf{1}^i)^{M^i}$ being already conditioned on $((\mathcal{I}^k_{M^k \cap M^i})^{M^i})_{k \in N}$, and the last equality follows from the law of iterated expectations. Therefore, together with the linearity of the expectations operator we have
\begin{eqnarray*}
	\mathbb{E}\left[\left((\tilde{Y}^{(1)})^{M^i} - (\tilde{Y}^{(2)})^{M^i}\right)(\mathbf{1}^i)^{M^i}\right] = \mathbb{E}\left[\left(Y^i_{M^i} - (\tilde{Y}^{(2)})^{M^i}\right)(\mathbf{1}^i)^{M^i}\right] = e^i.
\end{eqnarray*}
The seller's expected revenue is the sum of entry fees from all bidders plus the expected selling price in the second-price auction. The expected selling price in the second-price auction is the expectation of second highest order statistics across all bidders' estimated valuations from the seller's point of view. Therefore, the seller's expected revenue is $\sum_{i \in N} e^i + \mathbb{E}[(\tilde{Y}^{(2)})^M]$. We can further write out the expression as:
\begin{eqnarray*} & & \sum_{i \in N} e^i + \mathbb{E}[ (\tilde{Y}^{(2)})^M] \\
	& = & \mathbb{E}[(\tilde{Y}^{(1)})^M] + \sum_{i \in N} e^i - \left[\mathbb{E}[ (\tilde{Y}^{(1)})^M] - \mathbb{E}[ (\tilde{Y}^{(2)})^M]\right] \\
	& = & \mathbb{E}[(\tilde{Y}^{(1)})^M] + \sum_{i \in N} \mathbb{E}\left[\left((\tilde{Y}^{(1)})^{M^i} - (\tilde{Y}^{(2)})^{M^i}\right)(\mathbf{1}^i)^{M^i}\right] - \mathbb{E}\left[(\tilde{Y}^{(1)})^M - (\tilde{Y}^{(2)})^M\right] \\
	& = & \mathbb{E}[(\tilde{Y}^{(1)})^M] + \sum_{i \in N} \mathbb{E}\left[\left((\tilde{Y}^{(1)})^{M^i} - (\tilde{Y}^{(2)})^{M^i}\right)(\mathbf{1}^i)^{M^i}\right] - \mathbb{E}\left[\sum_{i \in N} ((\tilde{Y}^{(1)})^M - (\tilde{Y}^{(2)})^M) (\mathbf{1}^i)^{M}\right] \\
	& = & \mathbb{E}[(\tilde{Y}^{(1)})^M] + \sum_{i \in N} \mathbb{E}\left[\left((\tilde{Y}^{(1)})^{M^i} - (\tilde{Y}^{(2)})^{M^i}\right)(\mathbf{1}^i)^{M^i}\right] - \sum_{i \in N} \mathbb{E}\left[\left((\tilde{Y}^{(1)})^M - (\tilde{Y}^{(2)})^M\right) (\mathbf{1}^i)^{M}\right] \\
	& = & \mathbb{E}[(\tilde{Y}^{(1)})^M] + \sum_{i \in N} \left(e^i - (e^i)^M\right)
\end{eqnarray*} \hfill $\Box$

\subsection*{Proof of Corollary~\ref{Szech}}

If every bidder is aware of the same set of characteristics, $M^i = M' \subseteq M$ for all $i \in N$, then $M^k \cap M^i = M' = M^k \cap M$ for all $i, k \in N$, and hence $\left(\mathcal{I}^k_{M^k \cap M^i}\right)_{k \in N} = \left(\mathcal{I}^k_{M^k \cap M}\right)_{k \in N}$ for every $i \in N$. Hence, $\mathbb{E}[(\tilde{Y}^{(1)})^{M^i}] = \mathbb{E}[(\tilde{Y}^{(1)})^M]$ and $\mathbb{E}[(\tilde{Y}^{(2)})^{M^i}] = \mathbb{E}[(\tilde{Y}^{(2)})^M]$ for all $i \in N$. Therefore, the sum of the entry fees from the bidder can be written as
\begin{eqnarray*}
	\sum_{i \in N} e^i & = & \sum_{i \in N} \mathbb{E}\left[\left((\tilde{Y}^{(1)})^{M^i} - (\tilde{Y}^{(2)})^{M^i}\right)(\mathbf{1}^i)^{M^i}\right] \\
	& = & \sum_{i \in N} \mathbb{E}\left[\left((\tilde{Y}^{(1)})^{M} - (\tilde{Y}^{(2)})^{M}\right)(\mathbf{1}^i)^{M}\right] \\
	& = & \mathbb{E}\left[\sum_{i \in N}\left((\tilde{Y}^{(1)})^{M} - (\tilde{Y}^{(2)})^{M}\right)(\mathbf{1}^i)^{M}\right] \\
	& = & \mathbb{E} \left[(\tilde{Y}^{(1)})^{M} - (\tilde{Y}^{(2)})^{M}\right],
\end{eqnarray*}
where the first equality follows from Proposition~\ref{revenue}, the second equality is based on the condition that every bidder is aware of the same set of characteristics, and the third equality follows from the linearity of the expectations operator. Then the seller's expected revenue is $\mathbb{E}[(\tilde{Y}^{(1)})^{M} - (\tilde{Y}^{(2)})^{M}] + \mathbb{E}[(\tilde{Y}^{(2)})^{M}] = \mathbb{E} [(\tilde{Y}^{(1)})^{M}]$.\hfill $\Box$

\subsection*{Proof of Lemma~\ref{inequality}}

Let $W = \max\{\tilde{Y}^2_{M'}, \dots, \tilde{Y}^n_{M'}\}$. Since $\tilde{X}^1_{\ell}$ is independent of $\tilde{Y}^1_{M'}$ and $W$, conditioning on $(\tilde{Y}^1_{M'}, W)$ and applying Jensen's inequality to the convex function $x \mapsto \max\{\tilde{Y}^1_{M'} + x, W\}$ yields
\begin{eqnarray*} \mathbb{E}\left[\max\{\tilde{Y}^1_{M'} + \tilde{X}^1_{\ell}, W\} \mid \tilde{Y}^1_{M'}, W \right] & \geq & \max\{\tilde{Y}^1_{M'} + \mu^1_{\ell}, W\} \; \geq \; \max\{\tilde{Y}^1_{M'}, W\},
\end{eqnarray*} where the second inequality is strict on the event $\{\tilde{Y}^1_{M'} \geq W\}$ because $\mu^1_{\ell} > 0$. This event is $\{(\mathbf{1}^1)^{M} = 1\}$ and has positive probability by assumption. Taking expectations gives the claim. \hfill $\Box$

\subsection*{Proof of Lemma~\ref{unawareness_rents}}

For the term $\sum_{i \in N \setminus \{1\}} \left(e^i_z - (e^i_z)^M\right)$, every bidder $i \in N \setminus \{1\}$ believes that all the bidders are aware of the same set of characteristics, so the entry fee $e^i_z$ is computed based on this profile of information sets. However, the optimal entry fee from the point of view of an agent with full awareness $M$, $(e^i_z)^M$, considers the fact that bidder 1 is aware of characteristic $\ell$ in addition to the characteristics in $M'$. We claim that for all $i \neq 1$, 
\begin{eqnarray*} e^i_{z} & > & (e^i_z)^M \\
\mathbb{E}\left[((\tilde{Z}^{(1)})^{M^i} - (\tilde{Z}^{(2)})^{M^i})(\mathbf{1}^i_z)^{M^i}\right] & > & \mathbb{E}\left[((\tilde{Z}^{(1)})^{M} - (\tilde{Z}^{(2)})^{M})(\mathbf{1}^i_z)^{M}\right]
\end{eqnarray*} We distinguish two cases: First, conditional on $k \neq i$ winning the object, we have 
$$\mathbb{E}\left[((\tilde{Z}^{(1)})^{M^i} - (\tilde{Z}^{(2)})^{M^i})(\mathbf{1}^i_z)^{M^i} \mid k \mbox{ wins } \right] = \mathbb{E}\left[((\tilde{Z}^{(1)})^{M} - (\tilde{Z}^{(2)})^{M})(\mathbf{1}^i_z)^{M} \mid k \mbox{ wins} \right]$$ simply because both $(\mathbf{1}^i_z)^{M^i}$ and $(\mathbf{1}^i_z)^{M}$ must take the value of $0$ conditional on $k \neq i$ winning. 

Second, if $\mu_{\ell}^1 > 0$, then conditional on $i$ winning the object (recall that $i \neq 1$), an event that has positive probability by assumption, we have 
$$\mathbb{E}\left[((\tilde{Z}^{(1)})^{M^i} - (\tilde{Z}^{(2)})^{M^i})(\mathbf{1}^i_z)^{M^i} \mid i \mbox{ wins } \right] > \mathbb{E}\left[((\tilde{Z}^{(1)})^{M} - (\tilde{Z}^{(2)})^{M})(\mathbf{1}^i_z)^{M} \mid i \mbox{ wins} \right].$$ To see this, we distinguish two sub-cases. Conditional on $1$ not being the bidder with the second highest valuation, the expectations above are equal because the additional component that bidder $1$ is aware of is not relevant for either the first order or second order statistics. Conditional on $1$ being the bidder with the second highest valuation, we must have that the expectation of the second order statistic from the point of view of an agent with awareness $M^i$ is less than the one from the point of view of an agent with awareness $M$ because latter is aware of characteristic $\ell$ and $\mu_{\ell}^1 > 0$. This argument proves the claim. Therefore, if $\mu_{\ell}^1 > 0$, then we have $\sum_{i \in N \setminus \{1\}} \left(e^i_z - (e^i_z)^M\right) > 0$. \hfill $\Box$

\subsection*{Proof of Proposition~\ref{onemorebidder}}

Suppose the set of bidders in $\{1, 2, \dots, k\}$ with $k < n$ are aware of the set of characteristics $M'$ and characteristic $\ell \in M \setminus M'$ while the rest of the bidders are aware of $M'$ only. By Proposition~\ref{revenue}, the seller's expected revenue is $\mathbb{E}[(\tilde{Y}^{(1)})^M] + \sum_{i = k + 1}^n \left(e^i - (e^i)^M\right)$ where w.l.o.g. we can single out bidder $k + 1$'s difference in entry fees and rewrite the expected revenue as $$\mathbb{E}\left[(\tilde{Y}^{(1)})^M\right] + e^{k + 1} - (e^{k + 1})^M + \sum_{i = k + 2}^n \left(e^i - (e^i)^M\right).$$ 

Suppose now that bidder $k + 1$ becomes aware of characteristic $\ell$. The seller's expected revenue in this case are by Proposition~\ref{revenue}, $$\mathbb{E}[(\tilde{Z}^{(1)})^M] + \sum_{i = k + 2}^n \left(e^i_z - (e^i_z)^M\right).$$ 

Note that for all $i = k + 2, ..., n$, $e^i = e^i_z$. Thus, 
\begin{eqnarray*} \lefteqn{\mathbb{E}[(\tilde{Z}^{(1)})^M] + \sum_{i = k + 2}^n \left(e^i_z - (e^i_z)^M\right)} \\  & > & \mathbb{E}\left[(\tilde{Y}^{(1)})^M\right] + e^{k + 1} - (e^{k + 1})^M + \sum_{i = k + 2}^n \left(e^i - (e^i)^M\right)
\end{eqnarray*} is equivalent to 
\begin{eqnarray*} \mathbb{E}[(\tilde{Z}^{(1)})^M] - \mathbb{E}\left[(\tilde{Y}^{(1)})^M\right] & > & e^{k + 1} - (e^{k + 1})^M  + \sum_{i = k + 2}^n \left((e^i_z)^M - (e^i)^M\right)
\end{eqnarray*}\hfill $\Box$

\subsection*{Proof of Lemma~\ref{lhs1}}

The argument is that of Lemma~\ref{inequality} applied to bidder $k + 1$, with $W = \max\{\tilde{Y}^1_{M'} + \tilde{X}^1_{\ell}, \dots, \tilde{Y}^{k}_{M'} + \tilde{X}^{k}_{\ell}, \tilde{Y}^{k + 2}_{M'}, \dots, \tilde{Y}^{n}_{M'}\}$, which is unaffected by raising bidder $k + 1$'s awareness and independent of $(\tilde{Y}^{k+1}_{M'}, \tilde{X}^{k+1}_{\ell})$. \hfill $\Box$

\subsection*{Proof of Lemma~\ref{lhs2}} 

Let $M' \in \mathcal{M}$ with $M' \subsetneqq M$ and let $\ell \in M \setminus M'$. Suppose that bidders $\{1, ..., k\}$ are exactly aware of $M' \cup \{\ell\}$ while bidders $\{k+1, ..., n\}$ are exactly aware of $M'$ only, with $\ell \in M \setminus M'$. Assume that $\mu_{\ell}^{k + 1} > 0$.  When raising bidder $k + 1$'s awareness of $\ell$, then for any $i = k + 2, ..., n$, 
\begin{eqnarray*} (e^{i})^M & > & (e^{i}_z)^M  \\
\mathbb{E}\left[((\tilde{Y}^{(1)})^{M} -(\tilde{Y}^{(2)})^{M}) (\mathbf{1}^{i})^{M}\right]  & > & 
\mathbb{E}\left[((\tilde{Z}^{(1)})^{M} -(\tilde{Z}^{(2)})^{M}) (\mathbf{1}^{i}_z)^{M}\right] 
\end{eqnarray*} When $i$ wins in the weak dominant strategy equilibrium of the second-price auction, then $(\tilde{Y}^{(1)})^{M} = (\tilde{Y}^{i}_{M'})^M$. In this case, we also have \\ $(\tilde{Y}^{(2)})^{M} = \max\{\tilde{Y}^1_{M'} + \tilde{X}^1_{\ell}, ...,  \tilde{Y}^{k}_{M'} + \tilde{X}^{k}_{\ell}, \tilde{Y}^{k + 1}_{M'}, \tilde{Y}^{k + 2}_{M'}, ..., \tilde{Y}^{i - 1}_{M'}, \tilde{Y}^{i + 1}_{M'}, ... \tilde{Y}^{n}_{M'}\}$. Similarly, if $i$ wins in the weak dominant strategy equilibrium of the second-price auction after bidder $k + 1$ was made aware of characteristic $\ell$, then $(\tilde{Z}^{(1)})^{M} = (\tilde{Y}^{i}_{M'})^M$ and $(\tilde{Z}^{(2)})^{M} = \max\{\tilde{Y}^1_{M'} + \tilde{X}^1_{\ell}, ..., \tilde{Y}^{k}_{M'} + \tilde{X}^{k}_{\ell}, \tilde{Y}^{k + 1}_{M'} + \tilde{X}^{k + 1}_{\ell}, \tilde{Y}^{k + 2}_{M'}, ..., \tilde{Y}^{i - 1}_{M'}, \tilde{Y}^{i + 1}_{M'}, ... \tilde{Y}^{n}_{M'}\}$. Thus, above inequality becomes 
\begin{eqnarray*} \lefteqn{\mathbb{E}\left[((\tilde{Y}^i_{M'})^{M} -  \max\{\tilde{Y}^1_{M'} + \tilde{X}^1_{\ell}, ..., \tilde{Y}^{k}_{M'} + \tilde{X}^{k}_{\ell}, \tilde{Y}^{k + 1}_{M'}, \tilde{Y}^{k + 2}_{M'}, ..., \tilde{Y}^{i - 1}_{M'}, \tilde{Y}^{i + 1}_{M'}, ... \tilde{Y}^{n}_{M'}\})(\mathbf{1}^{i})^{M}\right]} \\ & > & \mathbb{E}\left[((\tilde{Y}^i_{M'})^{M} -  \max\{\tilde{Y}^1_{M'} + \tilde{X}^1_{\ell}, ..., \tilde{Y}^{k}_{M'} + \tilde{X}^{k}_{\ell}, \tilde{Y}^{k + 1}_{M'} + \tilde{X}^{k + 1}_{\ell}, \tilde{Y}^{k + 2}_{M'}, ..., \tilde{Y}^{i - 1}_{M'}, \tilde{Y}^{i + 1}_{M'}, ... \tilde{Y}^{n}_{M'}\})(\mathbf{1}^{i}_z)^{M}\right]
\end{eqnarray*} The inequality follows now from three facts: First, on the r.h.s. we have $\tilde{Y}^{k + 1}_{M'} + \tilde{X}^{k + 1}_{\ell}$ under the max function while we have only $\tilde{Y}^{k + 1}_{M'}$ at the l.h.s. Second, $\mu_\ell^{k + 1} > 0$. Third, the probability of bidder $i$ winning is lower when bidder $k + 1$ is aware of characteristic $\ell$ (r.h.s.) compared to when bidder $k + 1$ is only aware of $M'$ (l.h.s.) if $\mu_{\ell}^{k + 1} > 0$. \hfill $\Box$

\subsection*{Proof of Lemma~\ref{rhs1}} 

Let $M' \in \mathcal{M}$ with $M' \subsetneqq M$ and let $\ell \in M \setminus M'$. Suppose that bidders $\{1, ..., k\}$ are exactly aware of $M' \cup \{\ell\}$ while bidders $\{k+1, ..., n\}$ are exactly aware of $M'$ only, with $\ell \in M \setminus M'$. Assume that $\mu_{\ell}^i > 0$ for $i = 1, ..., k$.  
\begin{eqnarray*} e^{k + 1} & > & (e^{k + 1})^M \\
\mathbb{E}\left[((\tilde{Y}^{(1)})^{M'} -(\tilde{Y}^{(2)})^{M'}) (\mathbf{1}^{k + 1})^{M'}\right] & > & \mathbb{E}\left[((\tilde{Y}^{(1)})^{M} -(\tilde{Y}^{(2)})^{M}) (\mathbf{1}^{k + 1})^{M}\right] 
\end{eqnarray*} When bidder $k + 1$ wins in the weak dominant strategy equilibrium of the second-price auction, then $(\tilde{Y}^{(1)})^{M'} = \tilde{Y}^{k + 1}_{M'}$. In this case, we also have $(\tilde{Y}^{(2)})^{M'} = \max\{\tilde{Y}^1_{M'}, ..., \tilde{Y}^{k}_{M'}, \tilde{Y}^{k + 2}_{M'}, ..., \tilde{Y}^{n}_{M'}\}$. However, from an agent's perspective who aware of $M$, he knows that bidders $1$ to $k$ are aware of characteristic $\ell$. Thus, $(\tilde{Y}^{(2)})^{M} = \max\{\tilde{Y}^1_{M'} + \tilde{X}^1_{\ell}, ..., \tilde{Y}^{k}_{M'} + \tilde{X}^k_{\ell}, \tilde{Y}^{k + 2}_{M'}, ..., \tilde{Y}^{n}_{M'}\}$. Hence, above inequality can be rewritten as
\begin{eqnarray*} \lefteqn{\mathbb{E}\left[(\tilde{Y}^{k + 1}_{M'} - \max\{\tilde{Y}^1_{M'}, ..., \tilde{Y}^{k}_{M'}, \tilde{Y}^{k + 2}_{M'}, ..., \tilde{Y}^{n}_{M'}\}) (\mathbf{1}^{k + 1})^{M'} \right] } \\ & > &  \mathbb{E}\left[(\tilde{Y}^{k + 1}_{M'} - \max\{\tilde{Y}^1_{M'} + \tilde{X}^1_{\ell}, ..., \tilde{Y}^{k}_{M'} + \tilde{X}^k_{\ell}, \tilde{Y}^{k + 2}_{M'}, ..., \tilde{Y}^{n}_{M'}\}) (\mathbf{1}^{k + 1})^{M} \right]
\end{eqnarray*} If $\mu_{\ell}^i > 0$ for $i = 1, ..., k$, then $\max\{\tilde{Y}^1_{M'} + \tilde{X}^1_{\ell}, ..., \tilde{Y}^{k}_{M'} + \tilde{X}^k_{\ell}, \tilde{Y}^{k + 2}_{M'}, ..., \tilde{Y}^{n}_{M'}\} > \max\{\tilde{Y}^1_{M'}, ..., \tilde{Y}^{k}_{M'}, \tilde{Y}^{k + 2}_{M'}, ..., \tilde{Y}^{n}_{M'}\}$ a.s. Moreover, an agent aware of $M$ also realizes that the probability of $k + 1$ winning is lower than $k + 1$ perceives if $\mu_{\ell}^i > 0$, $i = 1, ..., k$, because bidder $k + 1$ does not realize that bidders $1$ to $k$ are aware of $\ell$ with $\mu_{\ell}^i > 0$, $i = 1, ..., k$. This proves the inequality. \hfill $\Box$

\subsection*{Proof of Proposition~\ref{privacy}}

Suppose originally the bidders are commonly aware of the set of characteristics $M'$ with $\ell \notin M'$. By Corollary~\ref{Szech}, the seller's expected revenue is therefore $\mathbb{E}[(\tilde{Y}^{(1)})^M]$. By raising the common awareness of characteristic $\ell$ among all bidders without providing any information on it, each bidder's estimated valuation increases by $\mu_{\ell}$, so that the seller's expected revenue becomes $\mathbb{E}[\max\{\tilde{Y}^1 + \mu_{\ell}, \tilde{Y}^2 + \mu_{\ell}, \dots, \tilde{Y}^n + \mu_{\ell}\}] = \mathbb{E}[\max\{\tilde{Y}^1, \tilde{Y}^2, \dots, \tilde{Y}^n\}] + \mu_{\ell}$, which is strictly greater than $\mathbb{E}[(\tilde{Y}^{(1)})^M]$ if and only if $\mu_{\ell} > 0$. \hfill $\Box$

\subsection*{Proof of Proposition~\ref{unaware}}

Suppose originally bidders are informed on the set of characteristics $M'$ where $\ell \notin M'$. With the finest information the seller's expected revenue is exactly $\mathbb{E}[(Y^{(1)})^M]$ which is equal to $\mathbb{E}[\max\{\sum_{j \in M'} X^1_j, \sum_{j \in M'} X^2_j, \dots, \sum_{j \in M'} X^n_j\}]$. When all bidders are aware of the set of characteristics, $M' \cup \{\ell\}$, the best the seller can obtain is $\mathbb{E}[\max\{\sum_{j \in M'} X^1_j + X^1_{\ell}, \sum_{j \in M'} X^2_j + X^2_{\ell}, \dots, \sum_{j \in M'} X^n_j + X^n_{\ell}\}]$.
\begin{eqnarray*}
	& & \max\{\sum_{j \in M'} X^1_j + X^1_{\ell}, \sum_{j \in M'} X^2_j + X^2_{\ell}, \dots, \sum_{j \in M'} X^n_j + X^n_{\ell}\} \\
	& \leq & \max\{\sum_{j \in M'} X^1_j, \sum_{j \in M'} X^2_j, \dots, \sum_{j \in M'} X^n_j\} + \max\{X^1_{\ell}, X^2_{\ell}, \dots, X^n_{\ell}\}
\end{eqnarray*}
where the inequality follows from the subadditivity of the maximum. Taking expectations and using the monotonicity and the linearity of the expectations operator,
\begin{eqnarray*}
	& & \mathbb{E}\left[ \max\{\sum_{j \in M'} X^1_j + X^1_{\ell}, \sum_{j \in M'} X^2_j + X^2_{\ell}, \dots, \sum_{j \in M'} X^n_j + X^n_{\ell}\} \right] \\
	& \leq & \mathbb{E} \left[\max\{\sum_{j \in M'} X^1_j, \sum_{j \in M'} X^2_j, \dots, \sum_{j \in M'} X^n_j\} \right]+ \mathbb{E} \left[\max\{X^1_{\ell}, X^2_{\ell}, \dots, X^n_{\ell}\}\right] \\
	& < & \mathbb{E} \left[\max\{\sum_{j \in M'} X^1_j, \sum_{j \in M'} X^2_j, \dots, \sum_{j \in M'} X^n_j\} \right]
\end{eqnarray*} if $\mathbb{E}[\max\{X^1_{\ell}, X^2_{\ell}, \dots, X^n_{\ell}\}] < 0$.

For the necessity part, let $i^*$ denote the index of a bidder with the highest valuation when all bidders are exactly aware of $M'$, that is, $i^* \in \arg\max_{i \in N} \sum_{j \in M'} X^i_j$, with ties broken, let's assume w.l.o.g., by the lowest index. Then $i^*$ is measurable with respect to $(X^i_j)_{j \in M', i \in N}$ and hence independent of $(X^i_{\ell})_{i \in N}$. Pointwise,
\begin{eqnarray*} \max_{i \in N} \left\{\sum_{j \in M'} X^i_j + X^i_{\ell}\right\} & \geq & \sum_{j \in M'} X^{i^*}_j + X^{i^*}_{\ell} \; = \; \max_{i \in N} \sum_{j \in M'} X^i_j + X^{i^*}_{\ell}.
\end{eqnarray*} Since the $X^i_{\ell}$ are identically distributed across bidders and independent of $i^*$, the random variable $X^{i^*}_{\ell}$ has the same distribution as $X_{\ell}$, so $\mathbb{E}[X^{i^*}_{\ell}] = \mu_{\ell}$. Taking expectations and using the monotonicity and the linearity of the expectations operator,
\begin{eqnarray*} \mathbb{E}\left[\max_{i \in N} \left\{\sum_{j \in M'} X^i_j + X^i_{\ell}\right\}\right] - \mathbb{E}\left[\max_{i \in N} \sum_{j \in M'} X^i_j\right] & \geq & \mu_{\ell}.
\end{eqnarray*} Hence, if raising common awareness of $\ell$ strictly decreases the seller's expected revenue, then $\mu_{\ell} < 0$. \hfill $\Box$

\subsection*{Proof of Proposition~\ref{perfectinfo}}

If all bidders are aware of the same set of characteristics, following Corollary~\ref{Szech} the seller's expected revenue is equal to $\mathbb{E}[(\tilde{Y}^{(1)})^{M}]$. By Jensen's inequality and the convexity of the maximum, it follows that $\mathbb{E}[(Y^{(1)})^{M}] \geq \mathbb{E}[(\tilde{Y}^{(1)})^{M}]$.\hfill $\Box$

\end{document}